PriMera Scientific Publications

# Vibe Coding with Google AI Studio Case Study

**Christian Mancas[1]* and Cristina Mancas[2]**

[1]*Mathematics and Computer Science Department, Ovidius University at Constanta, Romania*

[2]*Gymnasium School No. 86, Bucharest, Romania*

***Corresponding Author:** Christian Mancas, Ovidius University, Bd. Mamaia 124, Constanta, CT, Romania.

## Abstract

This paper presents a case study on using Google AI Studio (version Pro) to vibe code a medium-complexity software application by a non-programmer who, moreover, needs it in another language than English. We prove that, on one hand, almost anybody can successfully vibe code, not only developers, even if you do not speak or understand English, and, on the other, that although such a top AI product can build a reasonable approximation of a fine database software application, vibe coding still needs thorough testing of it, as, even if the initial specification documentation is crystal clear and almost exhaustive, the resulting app may exhibit bugs and not that intelligent development decisions.

***Keywords:*** vibe coding; Google AI Studio; relational database design; software application development; software application testing

## Abbreviations

AI = Artificial Intelligence.
app(s) = software application(s).
db(s) = database(s).
dev = software development.
e.g. = for example.
GUI(s) = graphical user interface(s).
i.e. = that is.
KB = kilobytes.
rdb(s) = relational database(s).
RDM = Relational Data Model.
RDBMS(es) = Relational Database Management System(s).
spec doc = specification documentation.

## Introduction

As succinctly defined in [1], *vibe coding* "is an emergent programming paradigm where developers primarily write code by interacting with code-generating large language models rather than writing code directly". In fact, as this paper also proves, almost anybody can vibe-code, not only developers.

The 2nd author of this paper, a Professor for Primary School Studies, is a big enthusiast in teaching digital competencies to her students, aged 6 to 11, as well as a skilled user of several software applications (apps) and Artificial Intelligence (AI) tools. She graduated 6 years ago a M.Sc. program in Educational Management from the Psychology and Educational Sciences Faculty of the Bucharest University, with absolutely no software development (dev) background or experience. Whenever she needed an app, it was developed by the first author of this paper. Several times over the years she wanted him to teach her, at least, dev basics, but she could never find free time for it.

Recently, among others, to get new AI abilities, especially including vibe coding, she enrolled in and successfully completed the online Coursera Google AI course [2, 3], during which she built a portfolio of 20+ artifacts using Google AI, which culminated with vibe coding a custom AI solution for solving one of her and her colleagues real workplace challenge: the *ExcursiiScoala* (*SchoolTrips*) app for managing school trips.

This paper is a case study describing her experience with establishing the corresponding specification documentation (spec doc), vibe coding it with Google AI Studio [4], testing the resulting app, and correcting the app's bugs. This case study proves that, although Google AI Studio is still making significant mistakes, when well used it is able in the end to empower non-programmers in their quest to get the apps they need without expert help.

Moreover, remarkably, both the spec doc, Google AI Studio replies, and resulting app are in Romanian, so vibe coding can be done even if you do not know English, and probably in any other natural language, obtaining apps having graphical user interfaces (GUIs) in foreign languages.

## Related Work

The browser-based developer prototyping environment Google AI Studio is frequently studied, analyzed, and used as a core research and industry tool in both scholarly and commercial literature [5, 6]. The official portal for foundational papers behind the Gemini models powering Google AI Studio [7] currently provides 264 publications. There also exists a Build with Google AI Forum [8].

Vibe coding is successfully performed as well with Claude [9, 10], ChatGPT [11, 12] and a dozen other such AI tools (e.g., [13]).

We used a special vibe-coding flavor, where the spec doc was mainly written using math and Datalog formalisms, as reported in [14].

## Materials and Methods

Figure 1 shows the English version of the spec doc fetched to Google AI Studio (in a New Project) as obtained from Romanian with Google Translate.

The spec doc was mainly written by the second author of this paper, who is a skilled MS Access [17] apps user. The first one added only the following 4 technical specs from the Relational Data Model (RDM) theory [16-18] and MS Access technology, to shorten the spec doc and minimize the risk of automatically generated serious bugs: "All foreign keys must have referential integrity enforced". "Of course, the subforms must be linked to the forms through the corresponding foreign keys. The foreign keys must be decoded by replacing their numerical values with the values of the corresponding semantic keys". "The columns containing the autogenerated numeric values for the primary keys will be hidden"., and "The application will be monolithic, single-user, without sharing locks (exclusive access) and with automatic compaction upon each closure".

Hello! I would like you to build for me an MS Access application called ExcursiiScoala for managing school trips that we organize with our classes of students from the schools where we teach. Schools have a unique name within localities; e.g., "Gimnazială no. 86" in "Bucharest", "Sector 3" county. Localities have unique names within counties; e.g., "Sinaia" in "Prahova" county. Counties have unique names. We are only interested in the counties of Romania. A school has, in each calendar year, several classes with distinct names; e.g., in the school "Gimnazială no. 86" in "Bucharest", "Sector 3" county, there is a class called "P-C" (preparatory C) in each calendar year. We are only interested in classes starting from 2020 and up to the current year. There are several students enrolled in a class, minimum 10, maximum 27. For students, we are only interested in their first and last names (concatenated) and date of birth; e.g., "Bader Jad", born on 1-Sept.-2020. There cannot be two students with the same first and last name and date of birth. We are only interested in students born after 2013 and who are at least 6 years old at the time of enrollment. In any calendar year, a student can be enrolled in at most one class. A trip is organized on a specific day, with a specific cost per student, for a class, to a tourist attraction. Tourist attractions have unique names within localities; e.g., "Art Museum" in "Constanța", "Constanța" county. Any student in that class and only that class can register for a trip. The date of payment of the trip fee must be stored for each student. All the above data, except the payment date, are mandatory. All foreign keys must have referential integrity enforced. Both queries and reports are required for students who have not paid the trip fee for which they registered until at least one day before the trip, per trip and student, as well as totals per trip (number of participating students), class (trips, students), school year (classes, trips, students), school (school years, classes, trips, students), general (schools, school years, classes, trips, students), but also, orthogonally, per tourist attractions.

***Figure 1:*** English version obtained with Google Translate of the Romanian spec doc.

Furthermore, two other queries are needed: one to enroll all students who are not enrolled in any class in a class and one to copy students enrolled in a class to another class. The application can have a standard Access menu (with the corresponding form titled "Menu" and the green in its background replaced with dark navy blue) organized as follows: the root (called "Main application menu") contains the submenus "Data management", "Queries" and "Reports", plus a button to exit the application; "Data management" contains a button to open one of the forms attached to the tables, labeled with their names (e.g., "Students", "Schools", "Excursions", "Tourist destinations"), plus a button to return to the main menu. The forms are desired in datasheet format, allowing editing of the data in the respective tables. The counties form (titled "Tourist destinations") must have the localities form as a subform (so there is no need for a "Localities" button); similarly, the tourist destinations form must be a subform of the localities form, the class composition form must be a subform of the classes form, the classes form a subform of the schools form, and the participants form must be a subform of the excursions form. Of course, the subforms must be linked to the forms through the corresponding foreign keys. The foreign keys must be decoded by replacing their numerical values with the values of the corresponding semantic keys; e.g., in the Excursions form the "Tourist destination" column must be a drop-down list displaying, in order, the name of the destination, the locality, and the county, separated from each other by a comma followed by a space, and the "Class" column must also be a drop-down list displaying, in order, the year, the name of the class, the of the school, of the locality and then of the county, separated from each other in the same way (by a comma followed by a space). The columns containing the autogenerated numeric values for the primary keys will be hidden. The calendar years will be displayed with 4 digits. The navigation panel will be hidden. The query and report submenus must be structured on at least 2 other levels, with a button to open each query/report, plus one to return to the higher hierarchical level. The application will be monolithic, single-user, without sharing locks (exclusive access) and with automatic compaction upon each closure. Closure requests will have to be explicitly confirmed. Upon opening, the application will automatically open the form containing its menu.

***Figure 1:*** (Continued).

Please note that the spec doc does not provide details on the orthogonal query and report per tourist attractions: we were also curious about Google AI Studio inventiveness in this respect.

After working for some 7 minutes, Google AI Studio built and gave us access to the app presented in the next paper Section and replied, in Romanian too, with its description: the corresponding English version obtained with Google Translate is shown in Figure 2.

To enable several users to simultaneously test the app, we then published it [19], which took some 3 minutes.

Then, the first author of this paper developed, in some 4 hours, an MS Access app also based on the spec doc from Figure 1, to serve as an expert benchmark. Figure 3 shows its main menu.

The first author of this paper tested only the quality of the db table and SQL query design, as well as the backup .json export. The second one tested the functional behavior of the app. Testing needed, in total, 24+ work hours.

The ExcursiiScoala application in Microsoft Access style was built according to all the specified requirements and business rules:

Architecture and implemented features:

1. Access main menu (Switchboard):

o Form titled "Menu" with the classic green background replaced with dark navy blue (#0b1e36).

o Root titled "Main application menu", containing:

♣ "Data management" submenu (dedicated buttons for Students, Schools, Excursions, Tourist attractions and return button).

♣ "Queries" submenu structured hierarchically on 2 additional levels (Finance & Arrears, Totals & Summaries by levels, Action Queries).

♣ "Reports" submenu structured on 2 additional levels (Finance, School levels, Tourist impact).

♣ "Exit application" button with explicit confirmation and automatic compaction.

o The application automatically opens this menu on startup.

2. Forms and Hierarchical Subforms (Datasheet):

o Tourist Attractions: Master-form for Counties, with subform for Localities and sub-subform for Tourist Attractions.

o Schools: Master-form for Schools, with subform for Classes and sub-subform for Class Composition (Enrolled Students) with norm monitoring (10–27 students).

o Excursions: Master-form for Excursions, with subform for Participants (filtered strictly from students of the respective class) and record of payment date.

***Figure 2:*** Google AI Studio summary of the app generated according to the spec doc from Figure 1.

o Students: Complete Datasheet for students (concatenated last and first names, date of birth).

o Autogenerated primary keys (numeric IDs) are hidden in the data view.

o Calendar years are displayed with 4 digits (2020 – current year).

o Foreign keys are semantically decoded:

♣ Tourist attraction in Excursions: [Objective Name], [Locality], [County].

♣ Class in Excursions: [Year], [Class Name], [School], [Locality], [County].

♣ Locality in Schools and Attractions: [Locality], [County].

3. Referential integrity and strict validation rules:

o Schools have unique names per locality.

o Localities have unique names per county (Romanian counties only).

o Tourist attractions have unique names per locality.

o Distinct class name per school and year (classes starting in 2020).

o Class size: minimum 10, maximum 27 students.

o Students born after 2013 (> 31.12.2013), minimum age of 6 years at registration, uniqueness on (Last name, First name, Date of birth).

o In any calendar year, a student can be enrolled in at most one class.

o Only students of that class can register for a trip.

***Figure 2:*** (Continued).

4. SQL Queries and Access Reports:

Students with unpaid fees at least 1 day before the trip (per trip and student).

o Totals per trip (number of participants, income, arrears).

o Totals per class (trips, students).

o Totals per school year (classes, trips, students).

Totals per school (school years, classes, trips, students).

o General totals (schools, school years, classes, trips, students).

o Orthogonal totals per tourist attractions (number of trips, visiting students, distinct schools).

o Action query 1: Enroll all unassigned students in a class (with a limit of 27 students).

o Action query 2: Copy students from one class to another class (with year and capacity check).

o All queries provide switching between the datasheet and the SQL Jet/ACE code, and the reports provide A4 printable layout (Print Preview).

5. MS Access runtime environment:

o Single-user with Exclusive Access (no sharing locks).

o Automatic compaction and repair on every close.

o Explicit confirmation dialog on exit.

o Side navigation panel is hidden by default (with possibility to show/collapse from the ribbon).

***Figure 2:*** (Continued).

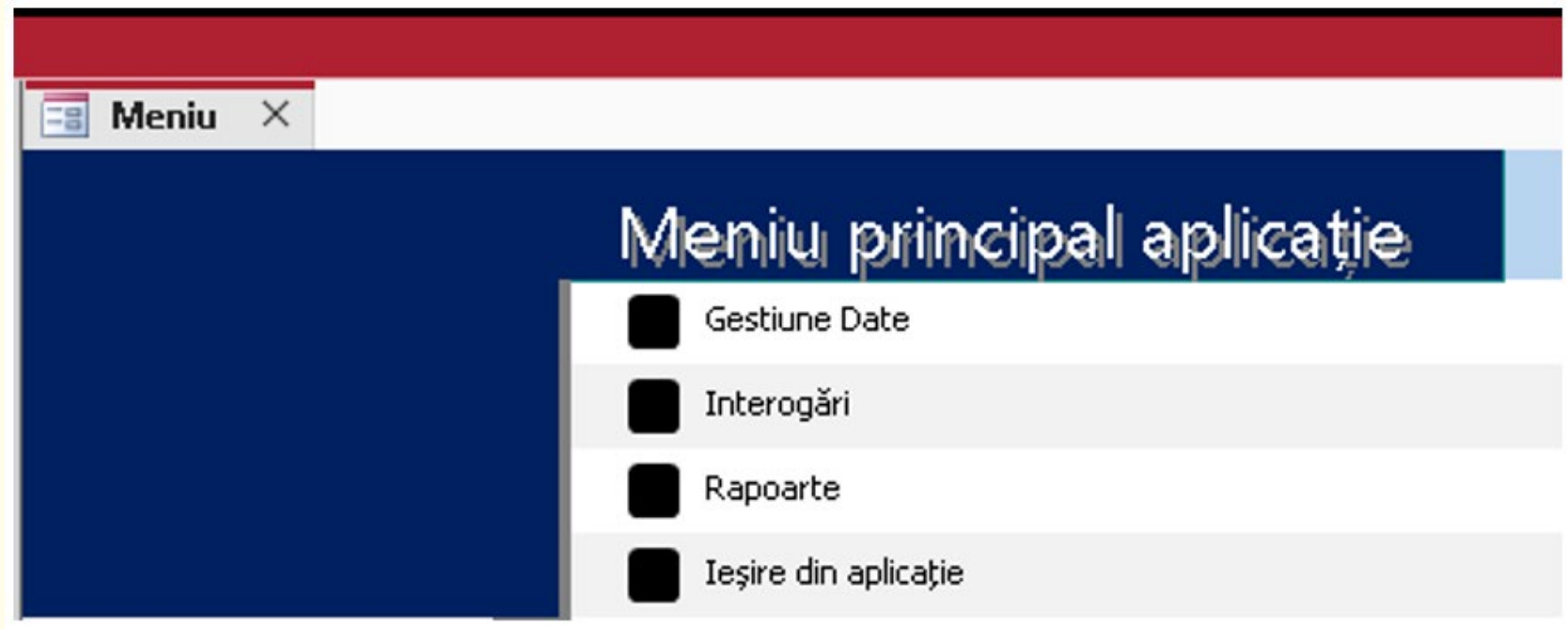

***Figure 3:*** Main menu of the benchmark MS Access app.

## Results and Discussion

### *The Main menu*

Figure 4 shows the first screen of the generated app, displaying its main menu.

At a first glance, it looks impressive, being web-based, having a finer GUI than the MS Access ones, and displaying everything in excellent Romanian, including the control-tip texts. From the uppermost bar, you can save and undo updates, learn the app's name and main features ("SchoolTrips : Database (File format Access 2016 - 2026) Microsoft Access [Exclusive Access - Monouser]"), open a *Navigation Pane* (which was not asked for), and close this webpage.

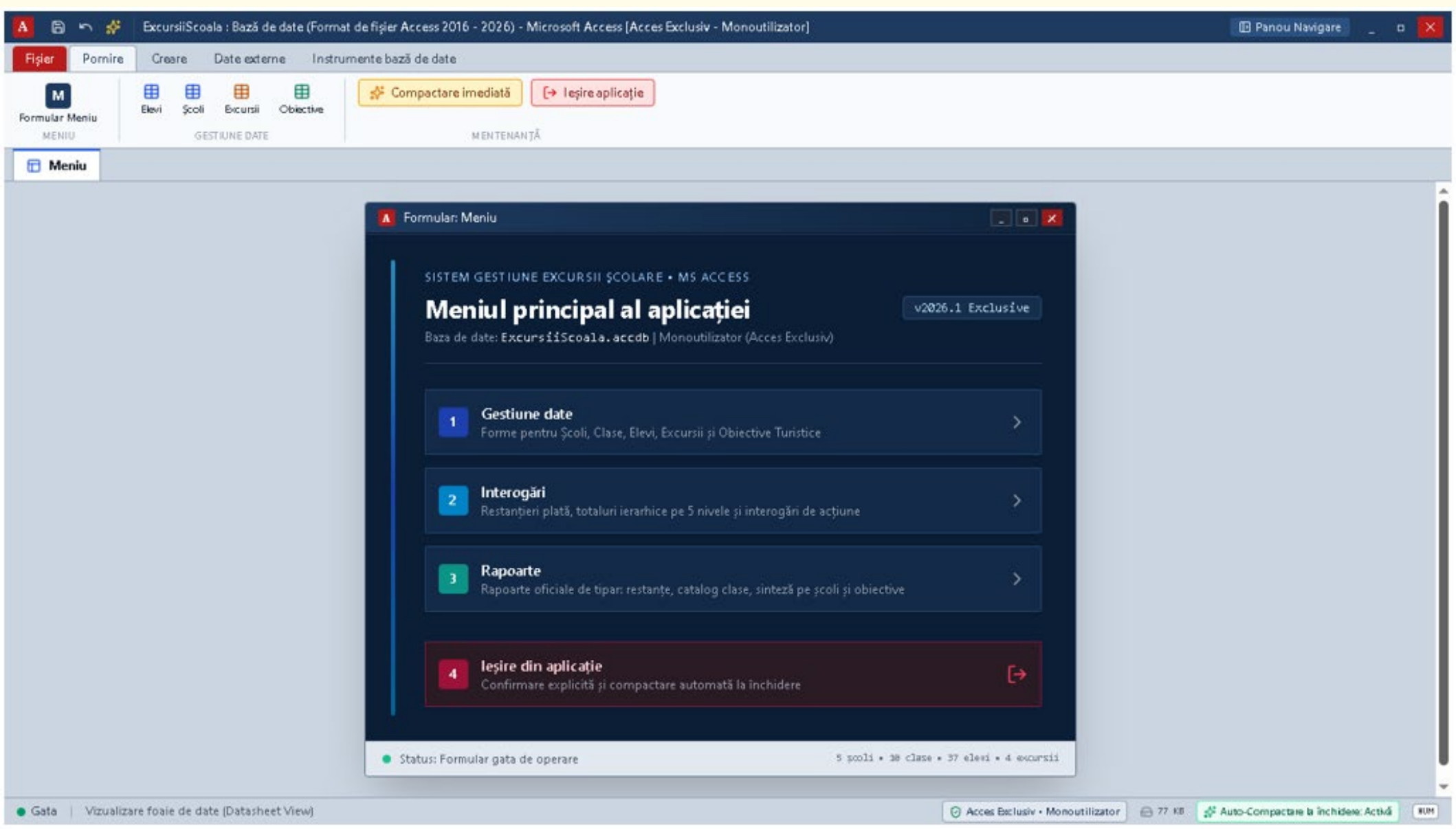

***Figure 4:*** First screen of the Google AI Studio generated app from the spec doc shown in Figure 1.

The *Navigation Pane* top content is shown in Figure 5. Exploring it fully, you learn that the underlying database (db) is made of 7 tables (*Students, Schools, Classes, ClassComposition, Trips, TripParticipants*, and *TouristicObjectives*), 5 forms (*Switchboard Menu, Students* (Datasheet), *Schools* (Master - Detail), *Trips* (Master - Detail), and *TouristicObjectives*), 10 queries, and 6 reports. The *Navigation Pane* is read-only, but when you click on any of its objects names the app opens that object. You can close it from either its X left-most upper button or the *Navigation Pane* one from the uppermost bar. Compared to the manually designed and implemented MS Access benchmark app, the generated one lacks two db tables: *Localities* and *Counties*.

The 2nd upper bar displays the names of 5 submenus that open on the 3rd one. The default one is the second, *Start* (*Pornire*), which contains the following 3 groups of buttons: *Menu* (containing only a read-only button reading *Menu Form*), *Manage Data* (containing 4 buttons, *Students*, *Schools*, *Trips*, and *Objectives*: clicking on any of it opens the corresponding form), and *Maintenance* (with two buttons: *Immediate Compacting* and *App Exit*). Opening any form comes with another pleasant surprise: the underlying db is populated with demo data, which contains actual Romanian schools, counties, cities, and touristic objectives!

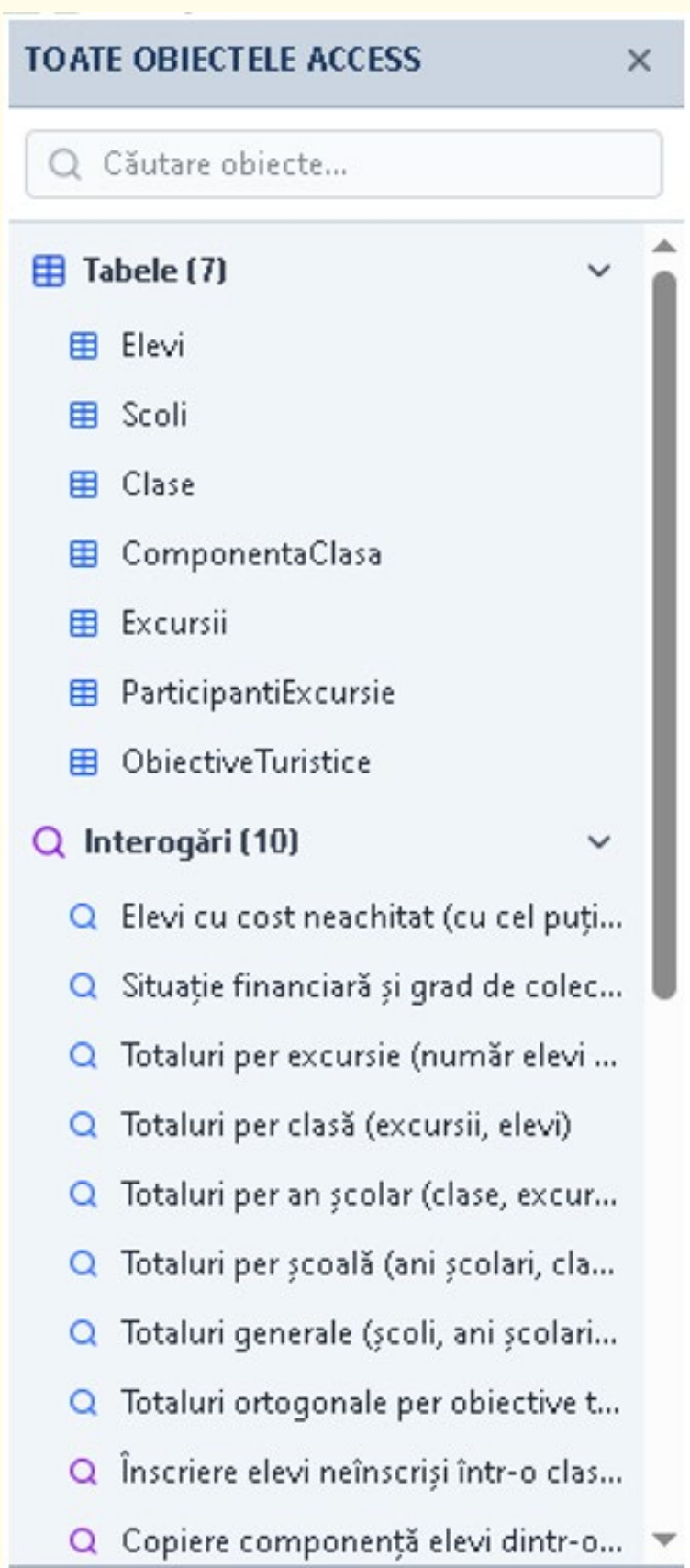

***Figure 5:*** App's Navigation Pane top content.

You can also exit the app by clicking on the fourth (red) option of the Main Menu shown in Figure 1. Exit is automatically followed by compacting the app's db and needs confirmation or can be canceled (by clicking on the *Anulare* (*Nu*) button), as shown in Figure 6.

The first submenu of this bar, *File* (*Fişier*), opens two unnamed groups: the first one is read-only and you learn from it the name of the underlying db (*ExcursiiScoala.accdb* in this case), its current size in KB, and the date and time of its last compaction; the second one contains two buttons: one for reinitializing the db demo instance and a fourth one for exiting the app.

The third submenu of this bar, *Create* (*Creare*), contains three buttons: *Debtors* (*Restanțieri*) and *General Totals* (*Totaluri Generale*) queries, and the *Debtors* report. Clicking on any of them opens the corresponding app object in a new tab.

For example, Figure 7 shows the *Debtors* query. In its uppermost bar you learn that it displays the result of running the SQL SELECT query computing the subset of students who did not pay the cost of the trips they enrolled in with at least one day before trip start. Unfortunately, neither student birth dates (mandatory for uniquely identifying students together with their names), nor localities and counties (mandatory for uniquely identifying both touristic objectives and schools, together with their names) are not displayed. In the *Payment Date* (*Data achitării*) column, null values are displayed as *Missing payment* date (*Lipsă data achitării*). A very useful *Payment Status* (*Stare achitare*) column was added, in which are displayed in red *Unpaid* (*Full debt*) and in black *Paid during* (or *after*) *trip day*. In fact, the query displays all students enrolled in trips, not only the debtor ones. Another pleasant surprise is that even the student names in their demo instance are plausible Romanian ones.

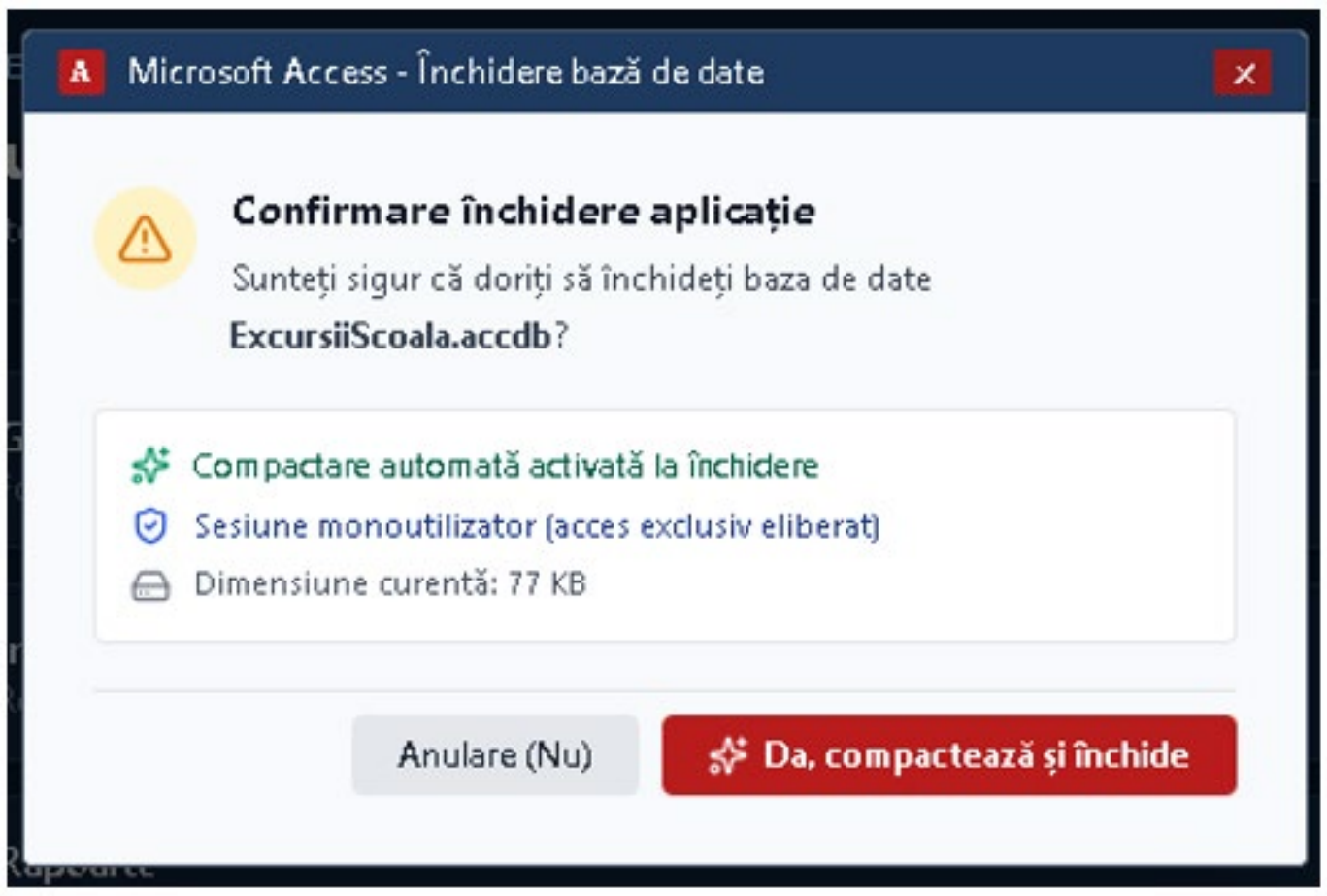

***Figure 6:*** App's exiting confirmation request form.

When clicking on the second, <> *Access SQL Query*, button, the current tab displays, also in read-only mode, the corresponding SQL code (which is completely irrelevant for non-db developers but correctly developed). Unfortunately, genuine full vibe coding is not possible, as developers may not alter or add SQL queries. For example, Figure 8 shows this code for the *Debtors* query.

Finally, the third button exports the current query data instance to a .csv file in the *Downloads* PC user system folder, which might be useful for both printing and MS Excel processing.

Figure 9 shows the top of the *Debtors* report. Unfortunately, it too lacks birth dates, localities, and counties data. In its uppermost bar, you can not only read its description and format but also print it (with the *Imprimare* button) and save its content (which includes only debtors) in your *Downloads* folder in .txt format (with the *Salvare* text button).

The fourth submenu, *External Data* (*Date externe*) contains only one button, *Export Backup Copy* (*Export copie de siguranță*), which saves in your *Download* folder a .json text file containing the db instance, from where you can infer that, however, the db schema also contains tables for storing localities and counties. Although a .accdb MS Access format is also advertised, it cannot be obtained.

Finally, the fifth submenu, *Database Tools* (*Instrumente baze de date*) contains two buttons: one for db compacting and repairing and one, *Relations Window* (*ERD Schema & Integrity*), for displaying the underlying db schema, shown in Figure 10.

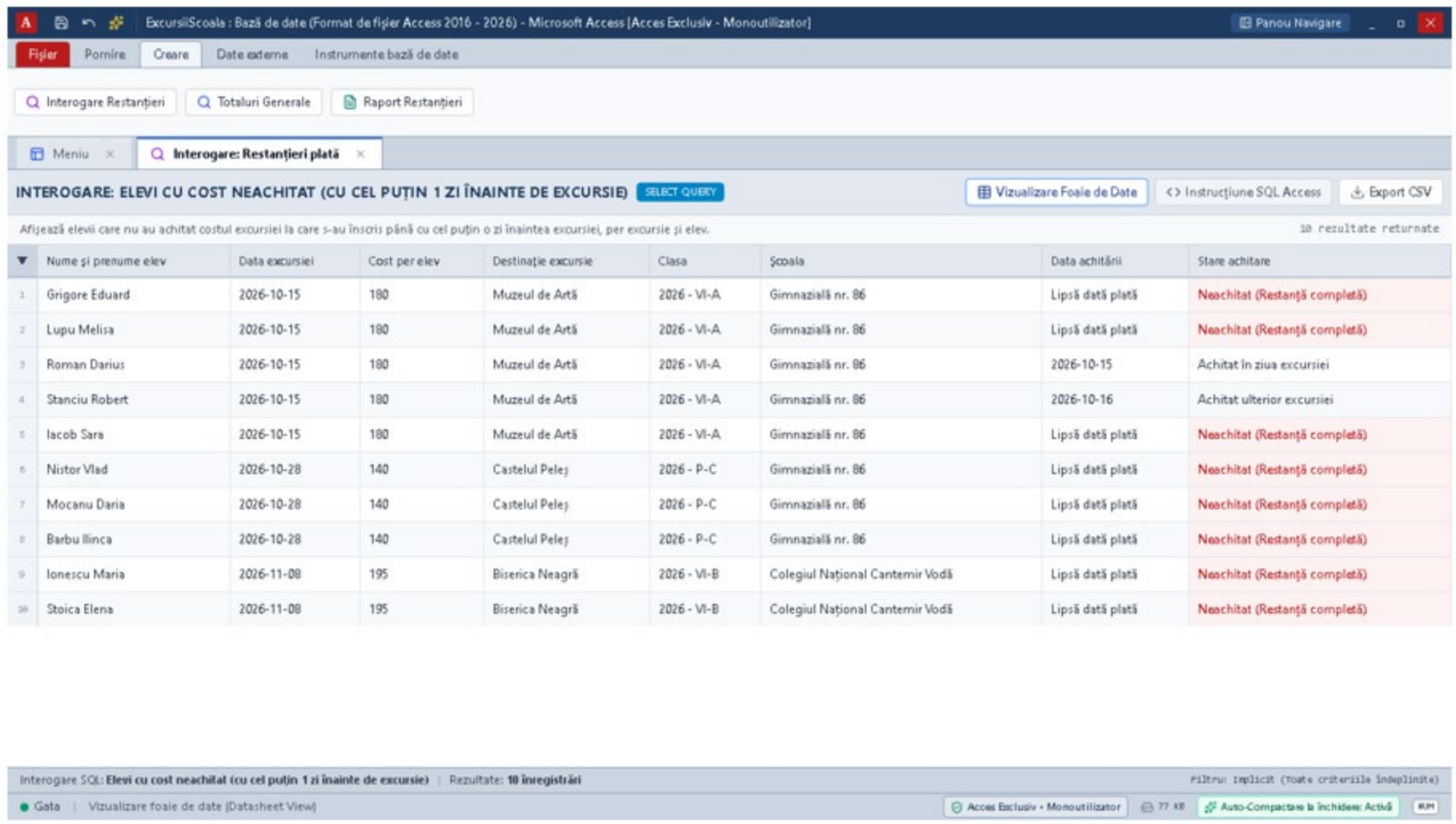

***Figure 7:*** Debtors query view.

```
Microsoft Access SQL Jet/ACE Engine View

SELECT E.NumeComplet AS [Elev], Ex.DataExcursiei, Ex.CostPerElev, Obj.Nume AS [Obiectiv], P.DataAchitarii,
       DateDiff('d', Nz(P.DataAchitarii, #01/01/1900#), Ex.DataExcursiei) AS [ZilePanaLaExcursie],
       IIf(P.DataAchitarii Is Null, 'Neachitat', 'Achitat cu întârziere / în ziua excursiei') AS [StarePlata]
FROM (((ParticipantiExcursie AS P
INNER JOIN Excursii AS Ex ON P.ExcursieId = Ex.Id)
INNER JOIN Elevi AS E ON P.ElevId = E.Id)
INNER JOIN ObiectiveTuristice AS Obj ON Ex.ObiectivId = Obj.Id)
WHERE (P.DataAchitarii Is Null)
   OR (P.DataAchitarii >= DateAdd('d', -1, Ex.DataExcursiei))
ORDER BY Ex.DataExcursiei, E.NumeComplet;
```

***Figure 8:*** Debtors query SQL code.

This confirms that the db has all 9 tables, including the ones for *Counties* (*Judete*) and *Localities* (*Localitati*), so the *Navigation Pane* is wrong about it. Fortunately, all of them have primary keys (very probably numerical with autonumbering, as they are not displayed in the corresponding forms) and all foreign keys are referencing them only.

Unfortunately, the foreign key names follow the not so intelligent *de facto* standard in this field, with the referenced table name suffixed by "Id", against which we proposed in vain for years (e.g., [18]) a much more intelligent one. For example, suppose that in the *Students* (*Elevi*) table we would need to add two columns for storing their birthplace and current living place, respectively. According to this *de facto* standard, as they should both be foreign keys referencing the *Localitati* table, they would be named *LocalitateId* and *LocalitateId1*! Much more intelligent is to name them *BirthPlace* and *LivingPlace*. Consequently, the "Id" suffix should be banned. For example, in table *ComponentaClasa* (displayed in magenta in Figure 10), *ClasaId* should be simply named *Clasa* (*Class*) and *ElevId* should be named *Elev* (*Pupil/Student*).

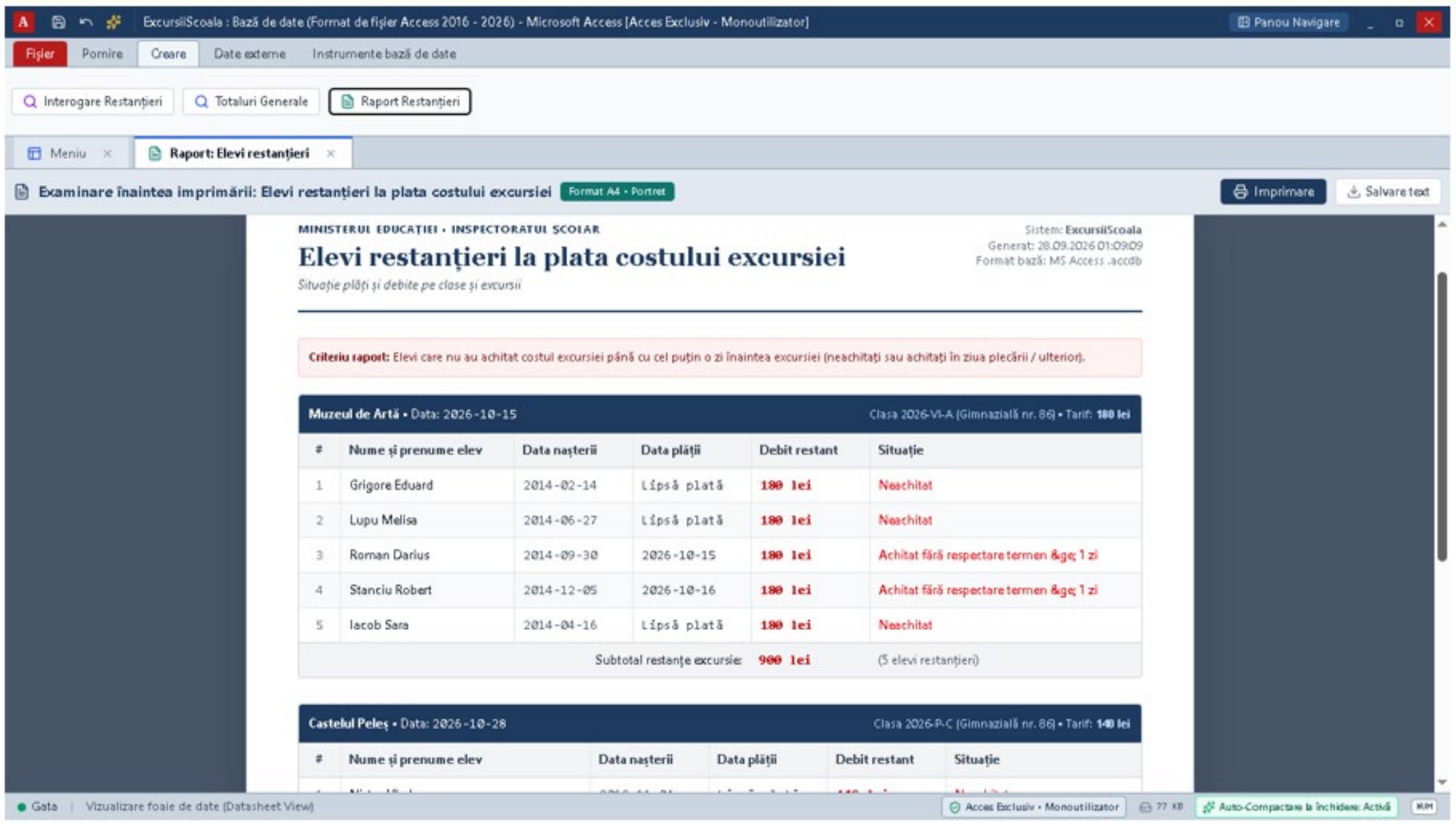

***Figure 9:*** Top of Debtors report.

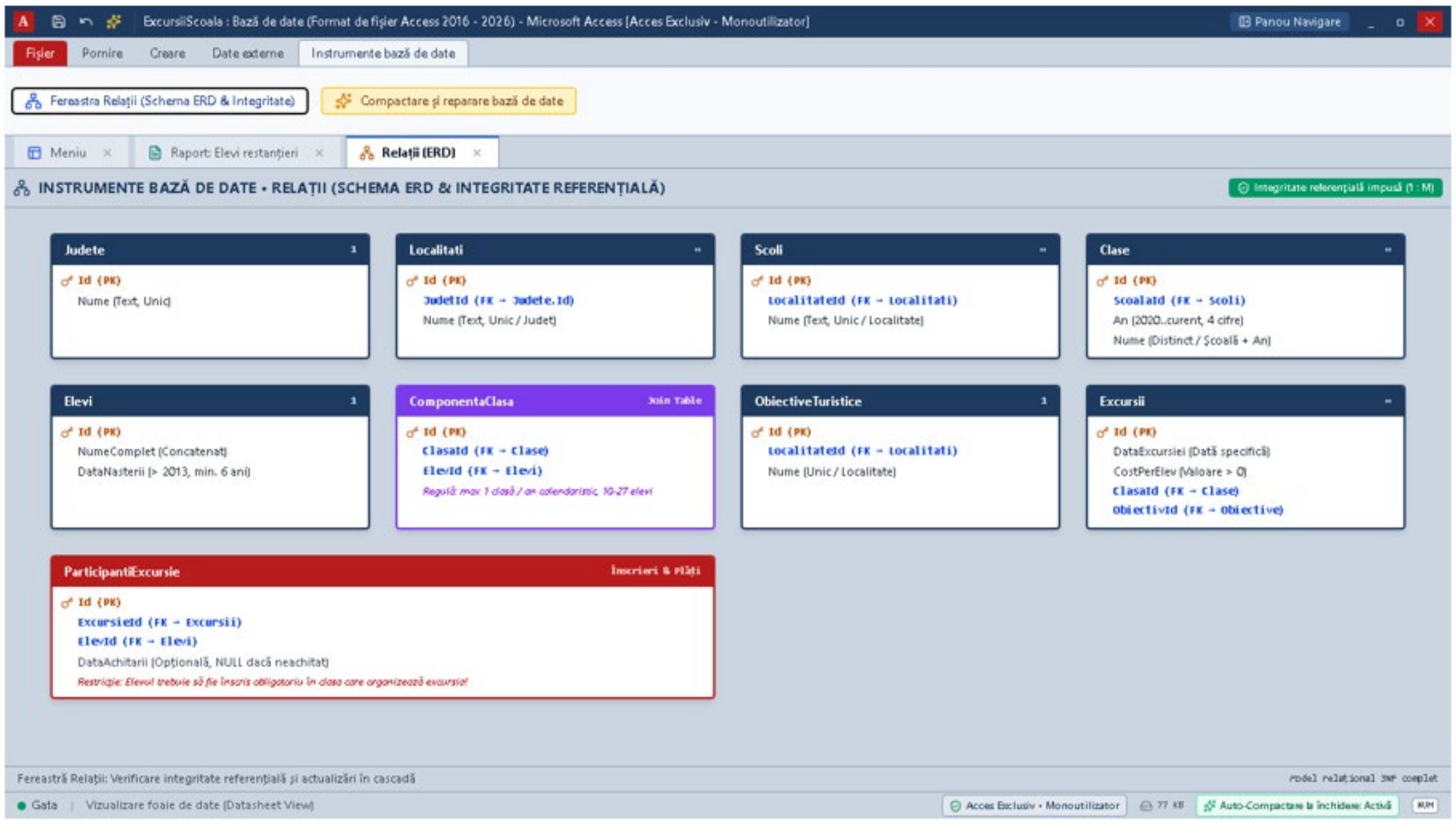

***Figure 10:*** The underlying db schema, including relational integrity.

### ***The Manage Data submenu***

Figure 11 shows the *Manage Data* (*Gestiune date*) submenu, which correctly contains buttons for opening all 4 forms asked for, as well as one for returning to the main menu (*Întoarcere în meniul principal*).

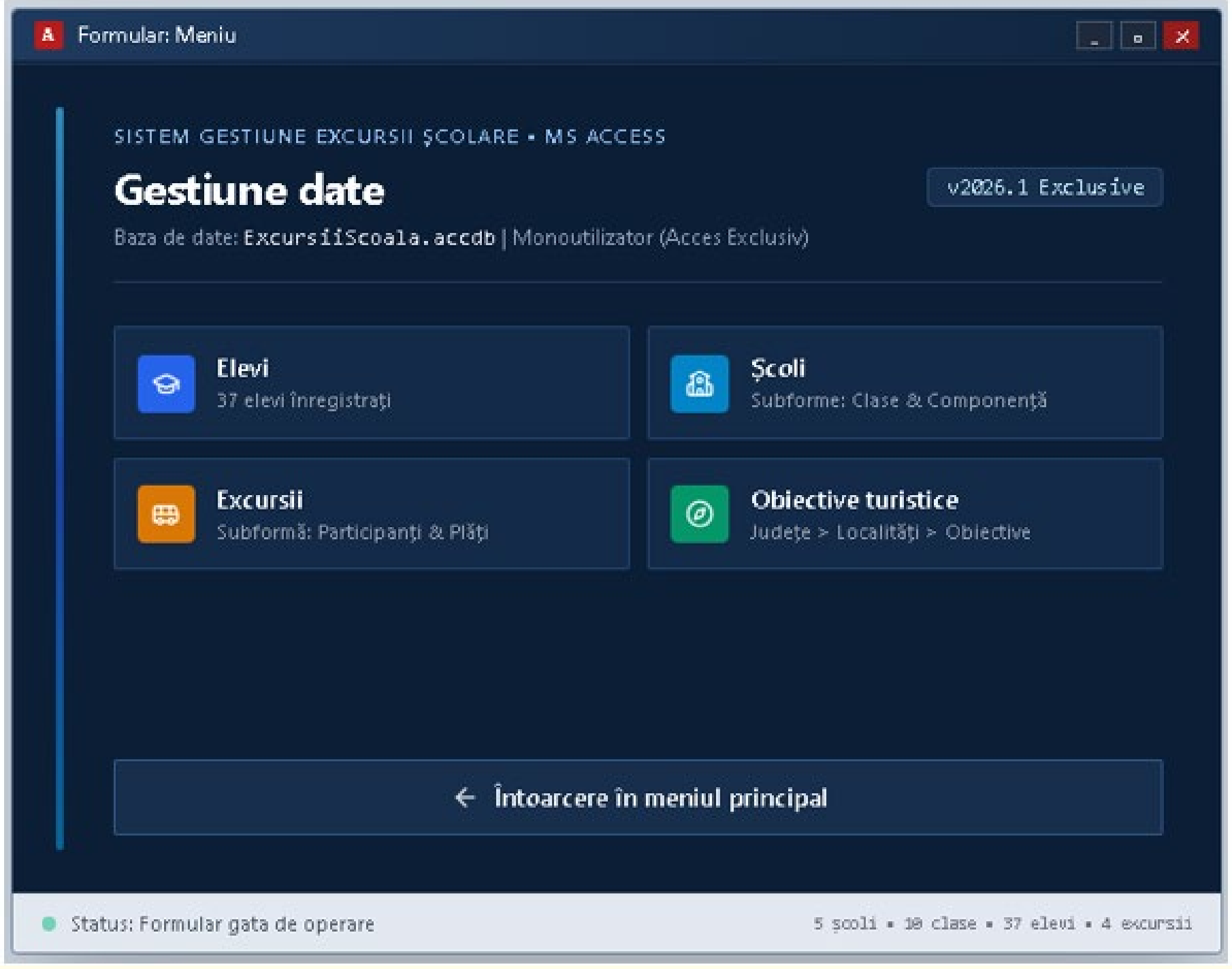

***Figure 11:*** The Manage Data submenu.

Figure 12 shows the *Elevi* (*Students/Pupils*) form. Unfortunately, unlike MS Access forms, you cannot resize column widths, change their positions between them, nor sort their instances by a column or column subset values, which is a minus for all other forms of this app. You cannot even copy a name or a birthdate value! Moreover, data is displayed unsorted, although it would be much easier for end-users to have it ascendingly sorted on names and then on birthdates.

The useful filtering control at the rightmost upper corner (*reading Căutare după nume sau an*) works fine to filter data by both birth year (*an*) and child name (*nume*), and not only for whole words, but also for any name substring. It is a pity that the other forms and subforms do not have such filtering capabilities, which should be added to reports as well.

Other useful unasked for additions are the computed columns *Age* (*Vârsta*) and *Current Class* (*Clasă curentă*). However, the latter one is confusing, as it displays all classes in which children have been enrolled in the course of time, in ascending chronological order, which is both redundant, as classes are listed for the current child once more in the rightmost form pane, and taking up too much space (think, e.g., of all currently displayed children being in their 13th class).

Finally, the rightmost pane, *Student Details* (*Fişă Elev*), displays redundantly the name, birth date, age, and current class, plus the history of class enrolments and trip participations of the currently selected student/pupil. Unfortunately, again, schools and touristic objective trip destinations may not be uniquely identified, as they lack corresponding localities and counties.

Fortunately, uniqueness of children is enforced: you cannot store two records with same name and birth date. Using the calendar control, you cannot set birth dates prior to 2013, but you can manually set and then store any date, including in the future, which violates two business rules: no student born before 2013 and no student less than 6 years old. Moreover, after updating birth dates, the computed column *Age* is not recomputed!

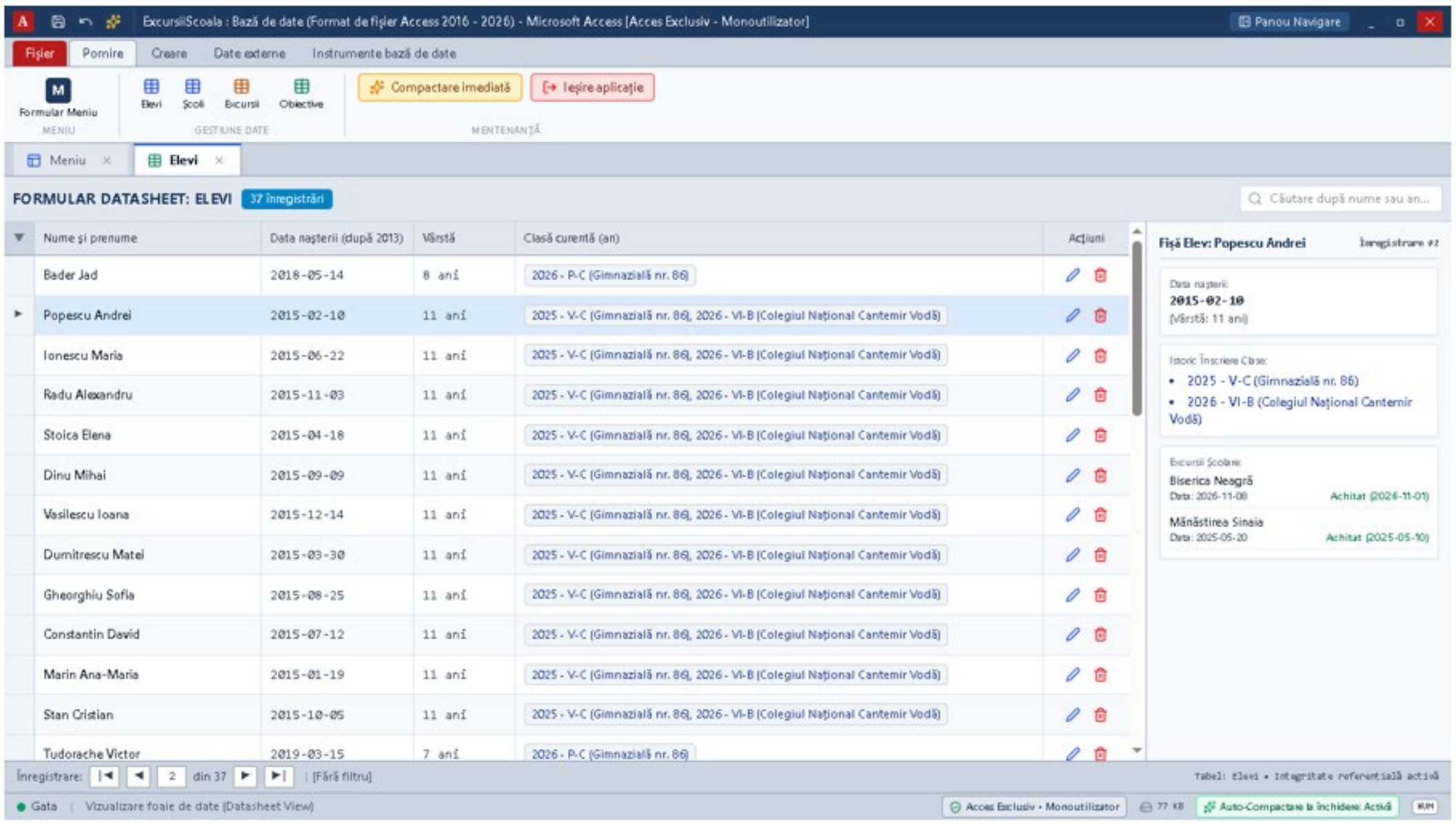

***Figure 12:*** The Students / Pupils form.

Figure 13 shows the *Schools* (*Școli*) form and its subforms. Please note first that, unlike MS Access that displays subforms vertically embedded, this app displays them horizontally, which becomes cumbersome when forms have dozen of columns and their embedding is done on dozens of levels. Again, data is displayed unsorted in all three panes.

Excellently, uniqueness is enforced in all three underlying db tables: you cannot store either two schools having same name in the same locality, or two classes having same name in the same year and school or enroll twice a children in a class.

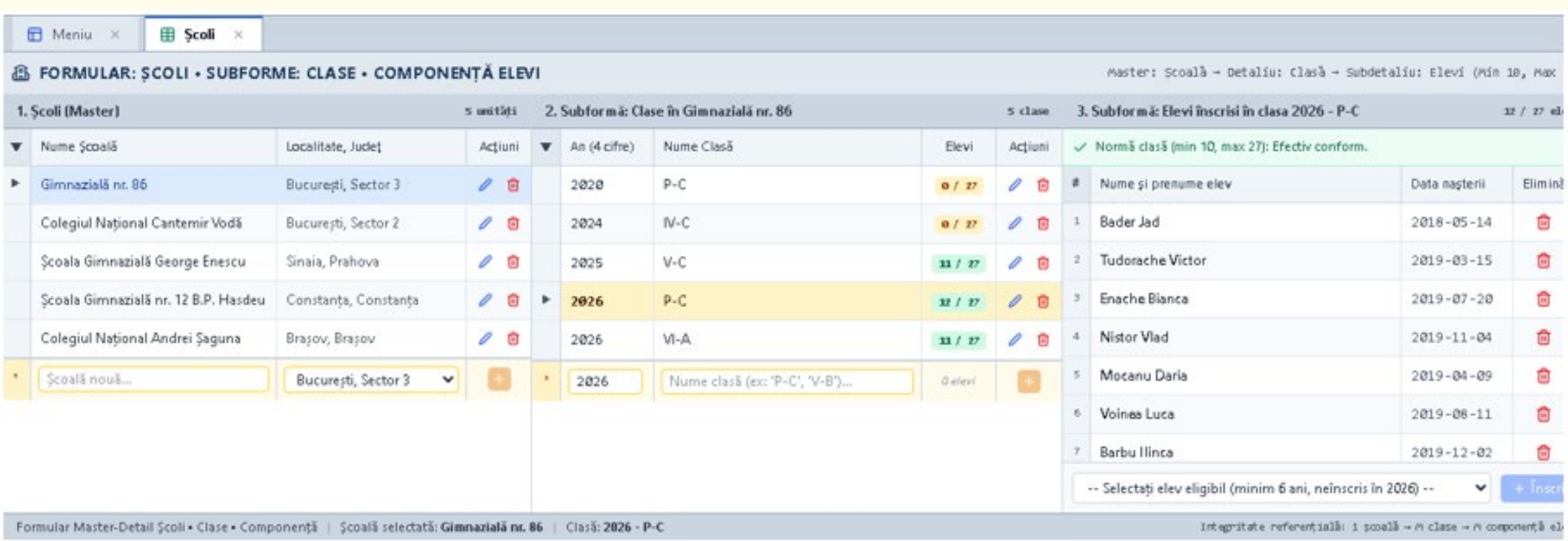

***Figure 13:*** The Schools form and its subforms.

Unfortunately, the instance of the combo-box *Locality*, *County* (*Județ*) is not alphabetically sorted, which makes end-user work extremely unpleasant. Another pleasant surprise is provided by the rightmost bottom combo-box for selecting children to be enrolled in the currently selected class: not only that its instance is alphabetically sorted but it contains only children that are not already enrolled that year and who are minimum six years old, as requested by the spec doc rules.

Figure 14 shows the *School Trips* (*Excursii*) form and its subform. Again, neither of its two instances are sorted.

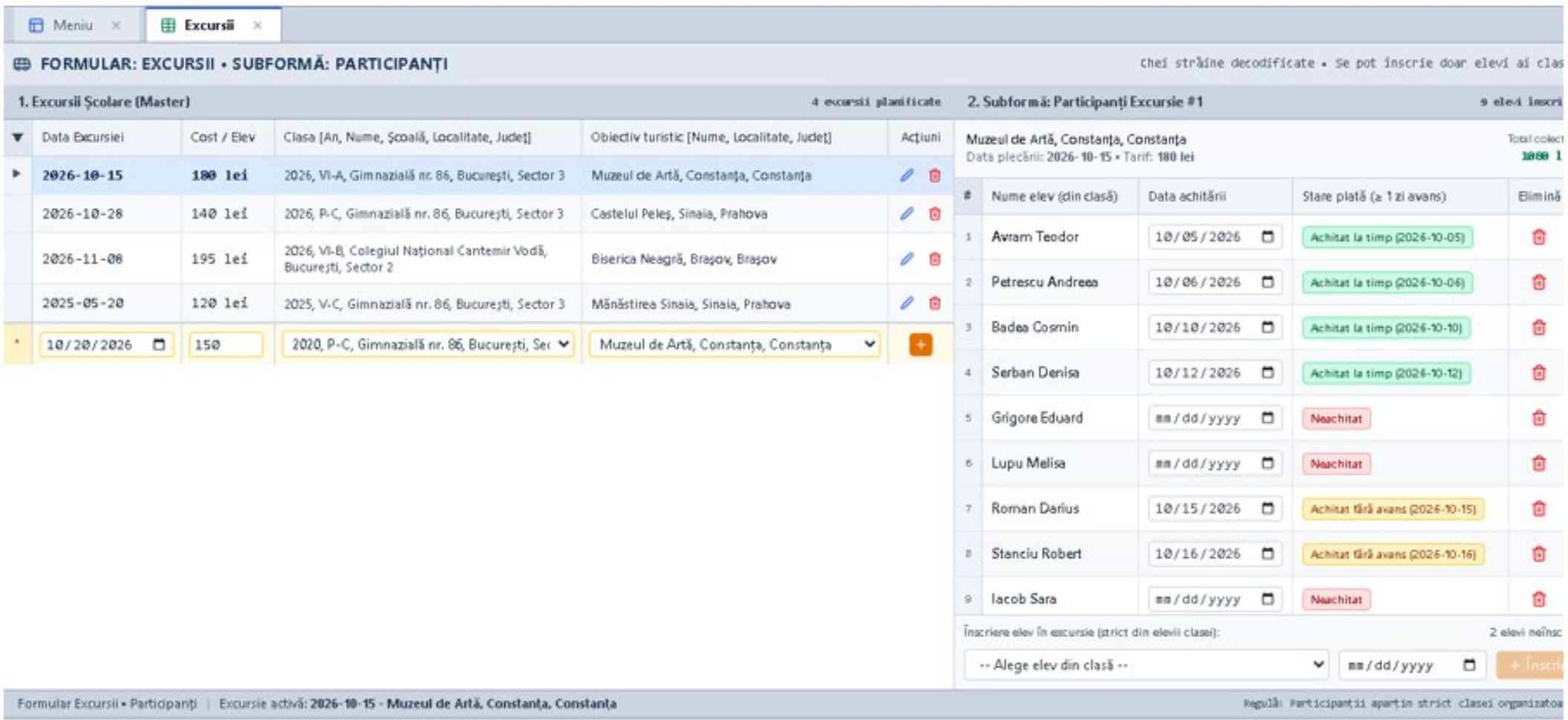

***Figure 14:*** The School Trips form and its subform.

Fortunately, like in the case of the *Enrolment* subform from Figure 13, you can add trip participants only from the students of the corresponding class who are not yet enrolled in the current trip.

Unfortunately, first, the instances of the combo-boxes *Class* and *Touristic Objective* are not sorted. Moreover, no semantic unique key has been enforced in the *Trips* table: we purposely did not include it in the spec doc, to see whether Google AI Studio knows that there may be no fundamental db table without at least one such key. Consequently, when testing, we could add two trips organized by the same class, to the same touristic objective, during the same day, which would not be serious in real life. According to [20], in such cases a composite semantic key made of all table columns except for the primary one must be added. In any similar contexts, this would be a superkey, as the *Cost* may be eliminated from it, but a superkey is better than no semantic key.

Moreover, a trivially inferable business rule (hence, not explicitly stated in the spec doc either) is not enforced: trips organized by a class may take place only during the class year or the next one (as the class year is the starting one, in the fall of that year, and ends in the summer of the next one). For example, we could successfully update the year of the trip to the Sinaia Monastery organized by class V-C of 2025 (last record from the left pane in Figure 14) to both 2030 (in the distant future!) and even in the first year of the current era, when Jesus was newborn!

Finally, Figure 15 shows, as requested, the *Touristic Objectives* (*Obiective turistice*) subform of the *Localities* (*Localități*), which is a subform of *Counties* (*Județe*) form. None of their instances is sorted either. Fortunately, all of them enforce their instance semantic uniqueness: you may not store either two counties with the same name, or two localities with the same name in the same county, or two objectives with the same name in the same locality.

***The Queries submenu***

Figure 16 shows the *Queries* (*Interogări*) submenu first level. Intelligently, Google AI Studio organized the requested queries into the following three groups: financial, totals, and action ones. Of course, it also added a button for returning to the main menu (*Întoarcere*...).

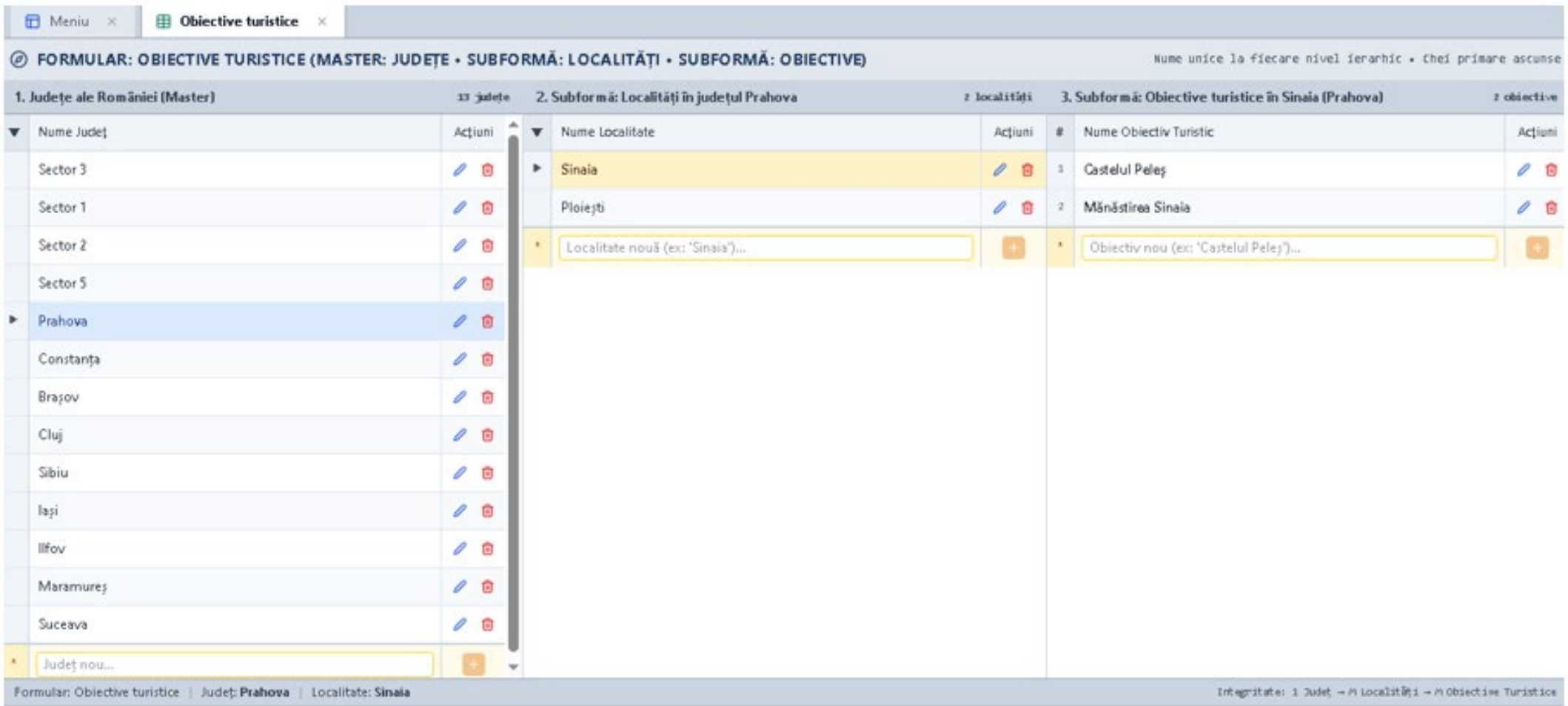

***Figure 15:*** The Romanian Counties form with its Localities and Touristic Objectives subforms.

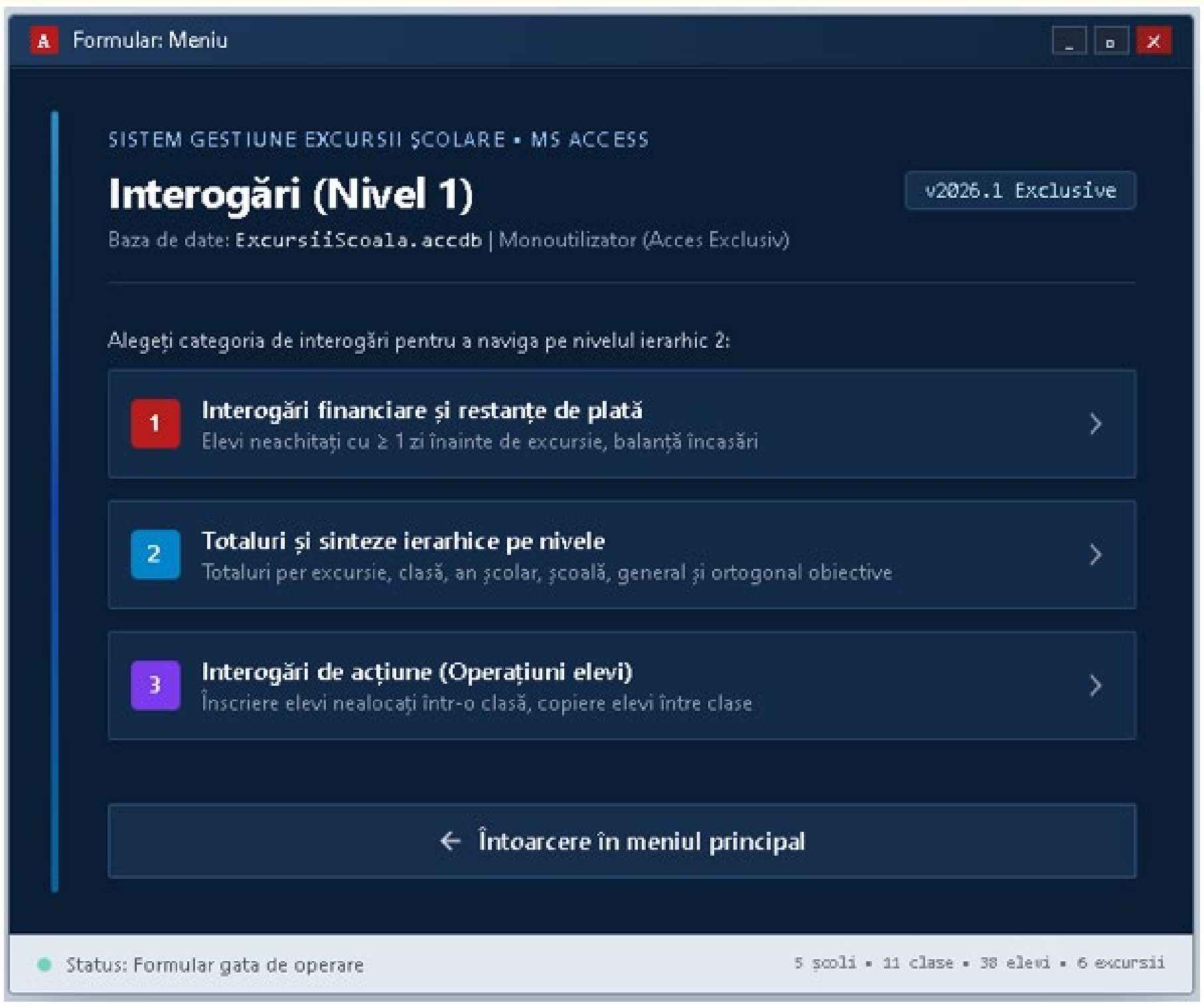

***Figure 16:*** The 3 query groups.

The two financial queries may be opened from the submenu shown in Figure 17. The first one opens the *Debtors* query shown in Figure 7. The second one, computing trip totals, is shown in Figure 18.

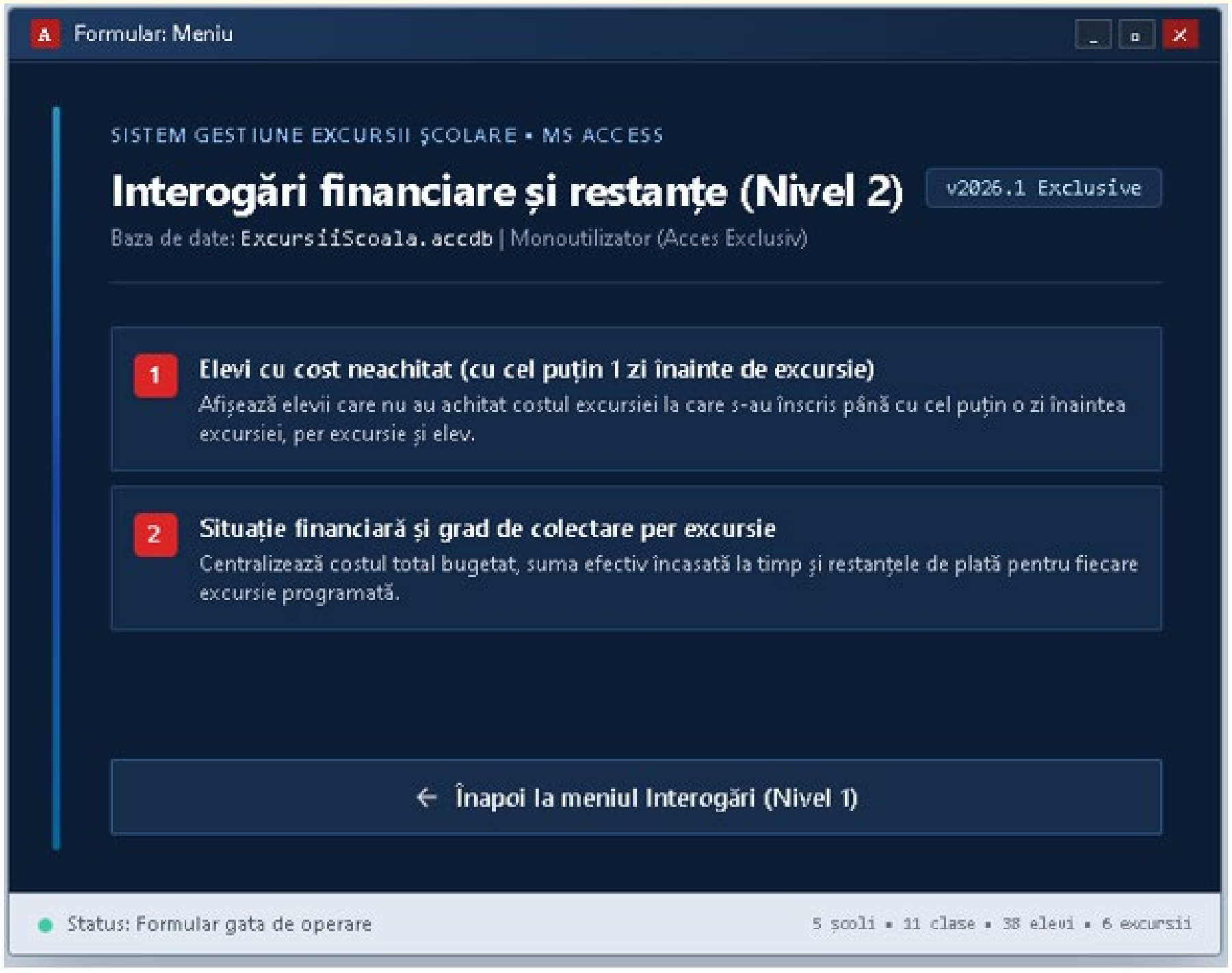

***Figure 17:*** The queries in the financial group.

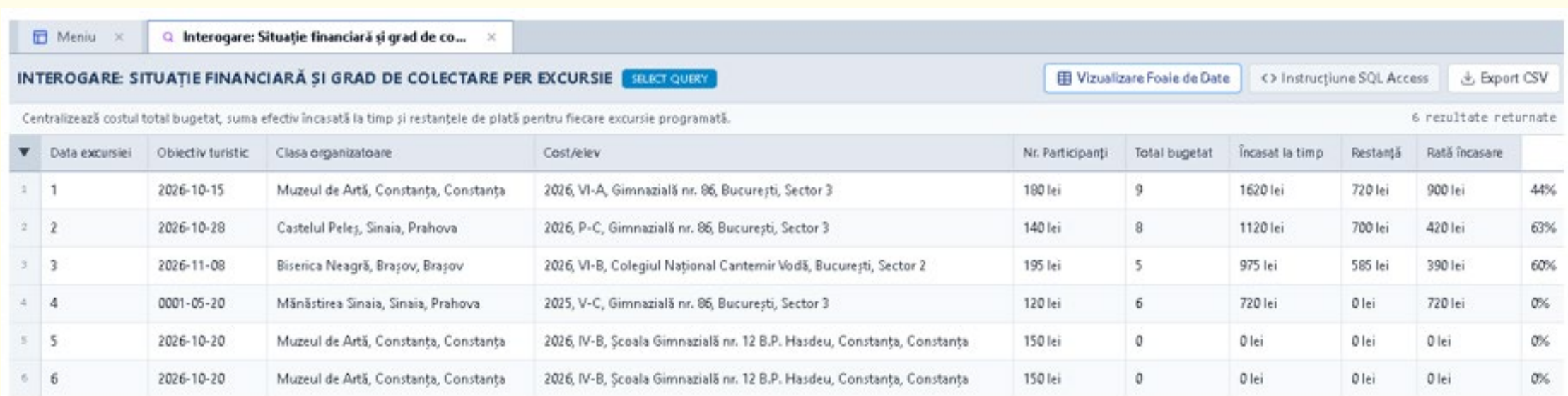

| ▼ | Data excursiei | Obiectiv turistic | Clasa organizatoare | Cost/elev | Nr. Participanți | Total bugetat | Încasat la timp | Restanță | Rată încasare | |
|---|---|---|---|---|---|---|---|---|---|---|
| 1 | 1 | 2026-10-15 | Muzeul de Artă, Constanța, Constanța | 2026, VI-A, Gimnazială nr. 86, București, Sector 3 | 180 lei | 9 | 1620 lei | 720 lei | 900 lei | 44% |
| 2 | 2 | 2026-10-28 | Castelul Peleș, Sinaia, Prahova | 2026, P-C, Gimnazială nr. 86, București, Sector 3 | 140 lei | 8 | 1120 lei | 700 lei | 420 lei | 63% |
| 3 | 3 | 2026-11-08 | Biserica Neagră, Brașov, Brașov | 2026, VI-B, Colegiul Național Cantemir Vodă, București, Sector 2 | 195 lei | 5 | 975 lei | 585 lei | 390 lei | 60% |
| 4 | 4 | 0001-05-20 | Mănăstirea Sinaia, Sinaia, Prahova | 2025, V-C, Gimnazială nr. 86, București, Sector 3 | 120 lei | 6 | 720 lei | 0 lei | 720 lei | 0% |
| 5 | 5 | 2026-10-20 | Muzeul de Artă, Constanța, Constanța | 2026, IV-B, Școala Gimnazială nr. 12 B.P. Hasdeu, Constanța, Constanța | 150 lei | 0 | 0 lei | 0 lei | 0 lei | 0% |
| 6 | 6 | 2026-10-20 | Muzeul de Artă, Constanța, Constanța | 2026, IV-B, Școala Gimnazială nr. 12 B.P. Hasdeu, Constanța, Constanța | 150 lei | 0 | 0 lei | 0 lei | 0 lei | 0% |

***Figure 18:*** The trip total query.

This query is wrong, as it includes as first column the trips id, as can be seen from its SQL code shown in Figure 19.

```
Microsoft Access SQL Jet/ACE Engine View

SELECT Ex.Id, Ex.DataExcursiei, Obj.Nume AS [Obiectiv], C.Nume AS [Clasa], Ex.CostPerElev,
       Count(P.Id) AS [NrParticipanti],
       Sum(Ex.CostPerElev) AS [TotalBugetat],
       Sum(IIf(P.DataAchitarii Is Not Null And P.DataAchitarii < Ex.DataExcursiei, Ex.CostPerElev, 0)) AS [IncasatLaTimp],
       Sum(IIf(P.DataAchitarii Is Null Or P.DataAchitarii >= Ex.DataExcursiei, Ex.CostPerElev, 0)) AS [Restant]
FROM (((Excursii AS Ex
INNER JOIN ObiectiveTuristice AS Obj ON Ex.ObiectivId = Obj.Id)
INNER JOIN Clase AS C ON Ex.ClasaId = C.Id)
LEFT JOIN ParticipantiExcursie AS P ON Ex.Id = P.ExcursieId)
GROUP BY Ex.Id, Ex.DataExcursiei, Obj.Nume, C.Nume, Ex.CostPerElev;
```

***Figure 19:*** The trip total query SQL SELECT statement.

Not only that this id is irrelevant for any end-user, but the instance of this query as shown in Figure 18 was computed with the SQL statement from Figure 19, while its header is misaligned, as the "Id" title is missing! Probably, in this case, the AI tool computed the header with a similar SQL statement from which Ex.Id was removed and hallucinated by faking its instance with the one computed by the query from Figure 19. The result is that the header is shifted one column position to the left, which would not be accepted by any end-user.

The 6 queries of the second group may be opened from the submenu shown in Figure 20.

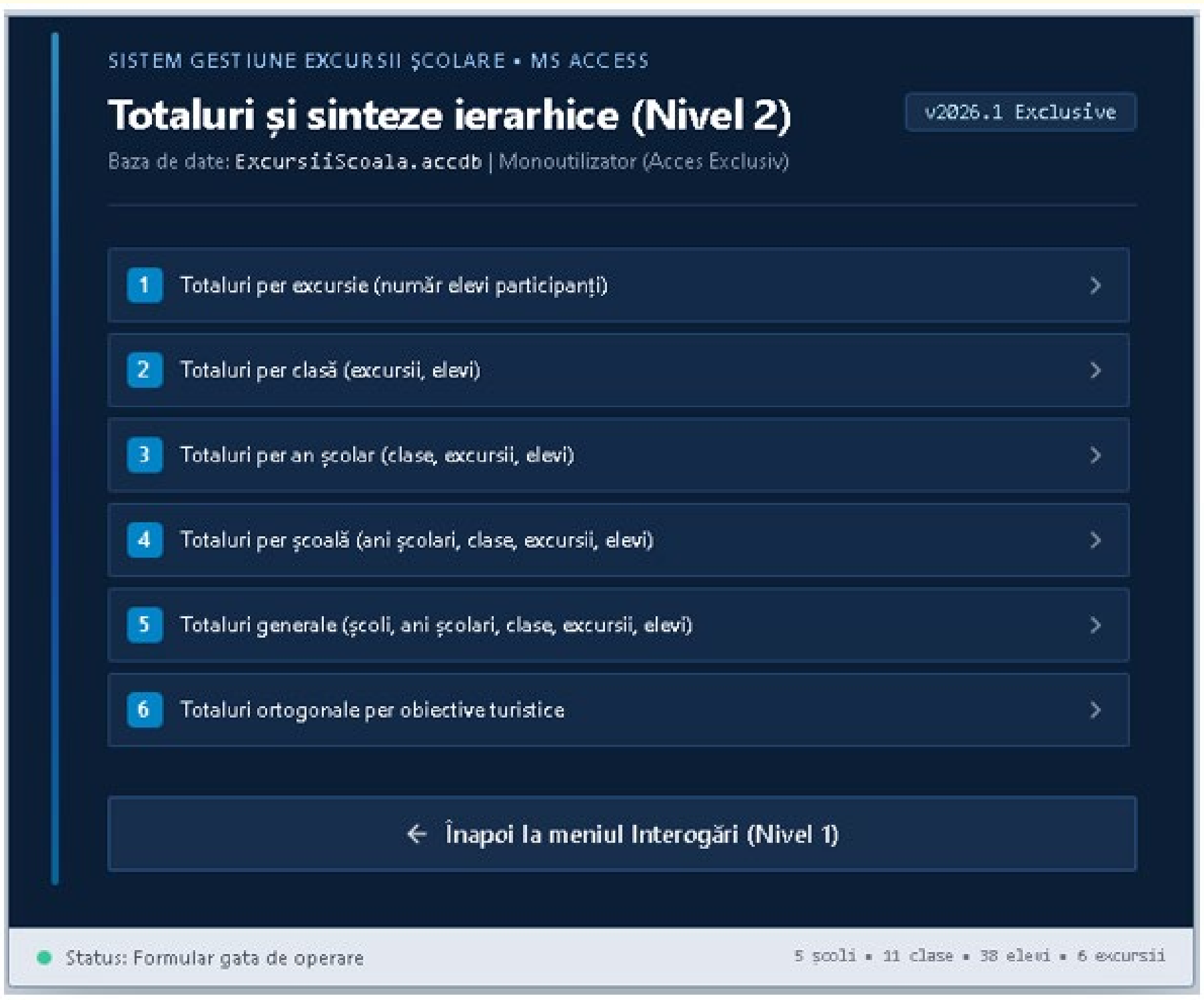

***Figure 20:*** The queries in the totals group.

The first one displays the same "hallucination" as the one from Figure 18. The second one lacks county and locality names for schools. The other 4 are ok. Unfortunately, each time you close one of these queries, the app is automatically redirecting you to the main menu.

Figure 21 shows the group of the action queries.

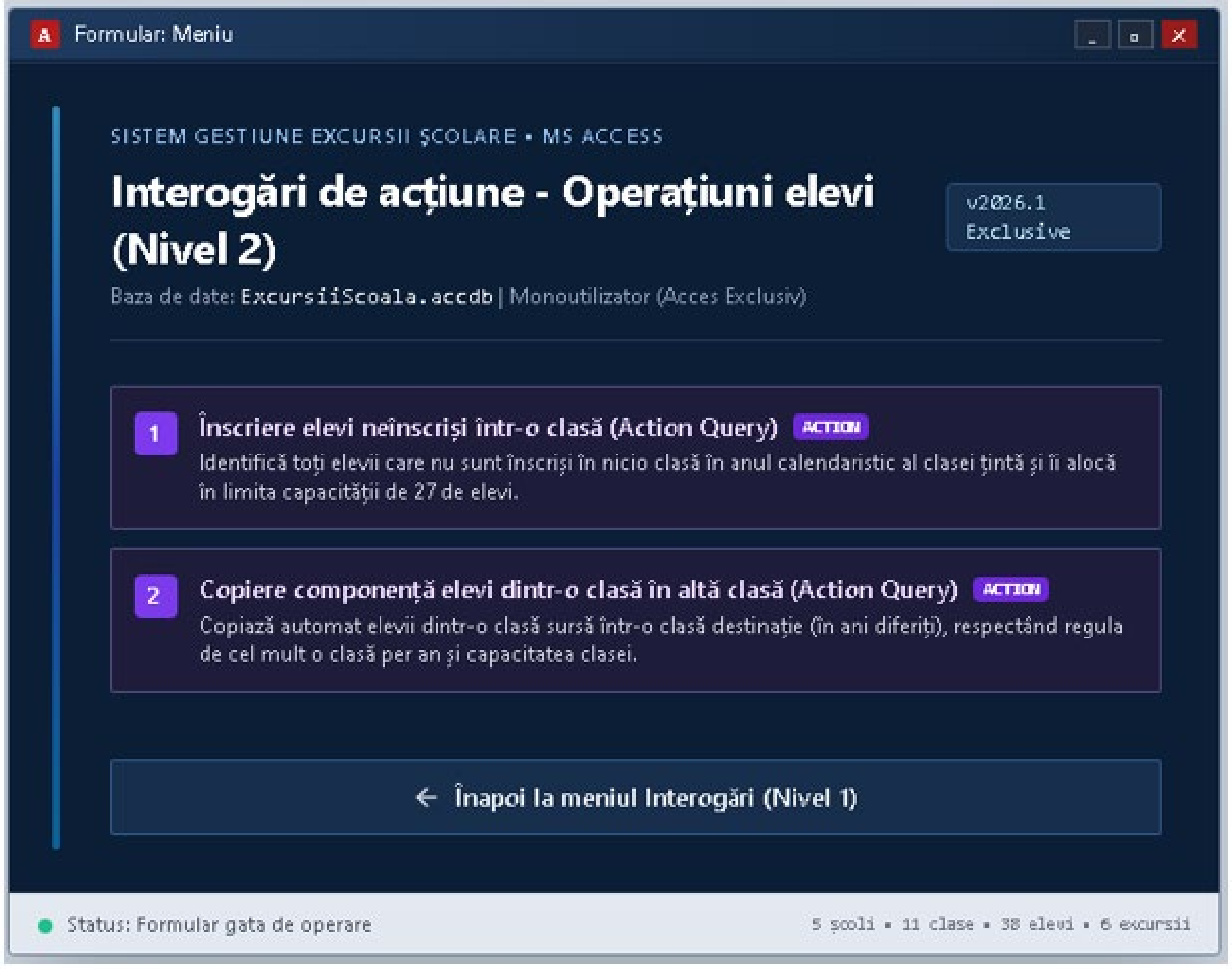

***Figure 21:*** The queries in the actions group.

Figure 22 shows the screen of the first action query, designed to bulk-enroll children in a desired class. Unfortunately, the instance of the combo-box (for selecting the destination classes for bulk children enrolment) is not sorted. Fortunately, the query works ok and, moreover, we also discovered while testing it that the combo-box instance is automatically recomputed when the set of classes is updated.

Figure 23 shows the screen of the second action query. Unfortunately, the instances of the two combo-boxes (for selecting the source and the destination classes for bulk children enrolment) are not sorted. Fortunately, the query works ok and both combo-boxes are automatically recomputed.

***The Reports submenu***

The Reports submenu (see Figure 24) is also organized on two levels, with reports carried out into the following 3 groups: financial, hierarchical, and touristic. Unfortunately, its 2nd level submenus display the same unpleasant behavior as the query ones: as soon as you close a report, you are automatically returned to the main menu.

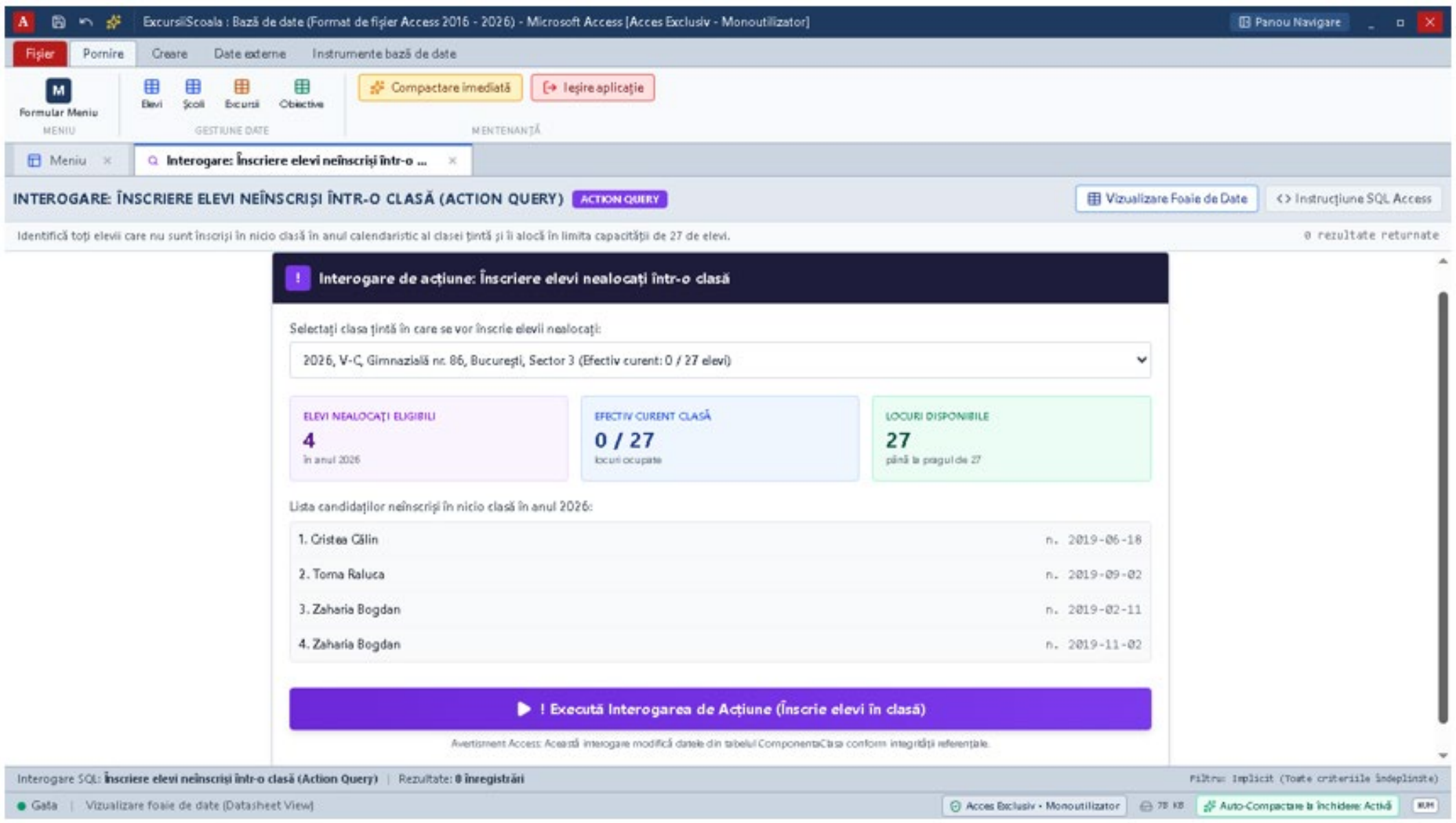

***Figure 22:*** The first action query.

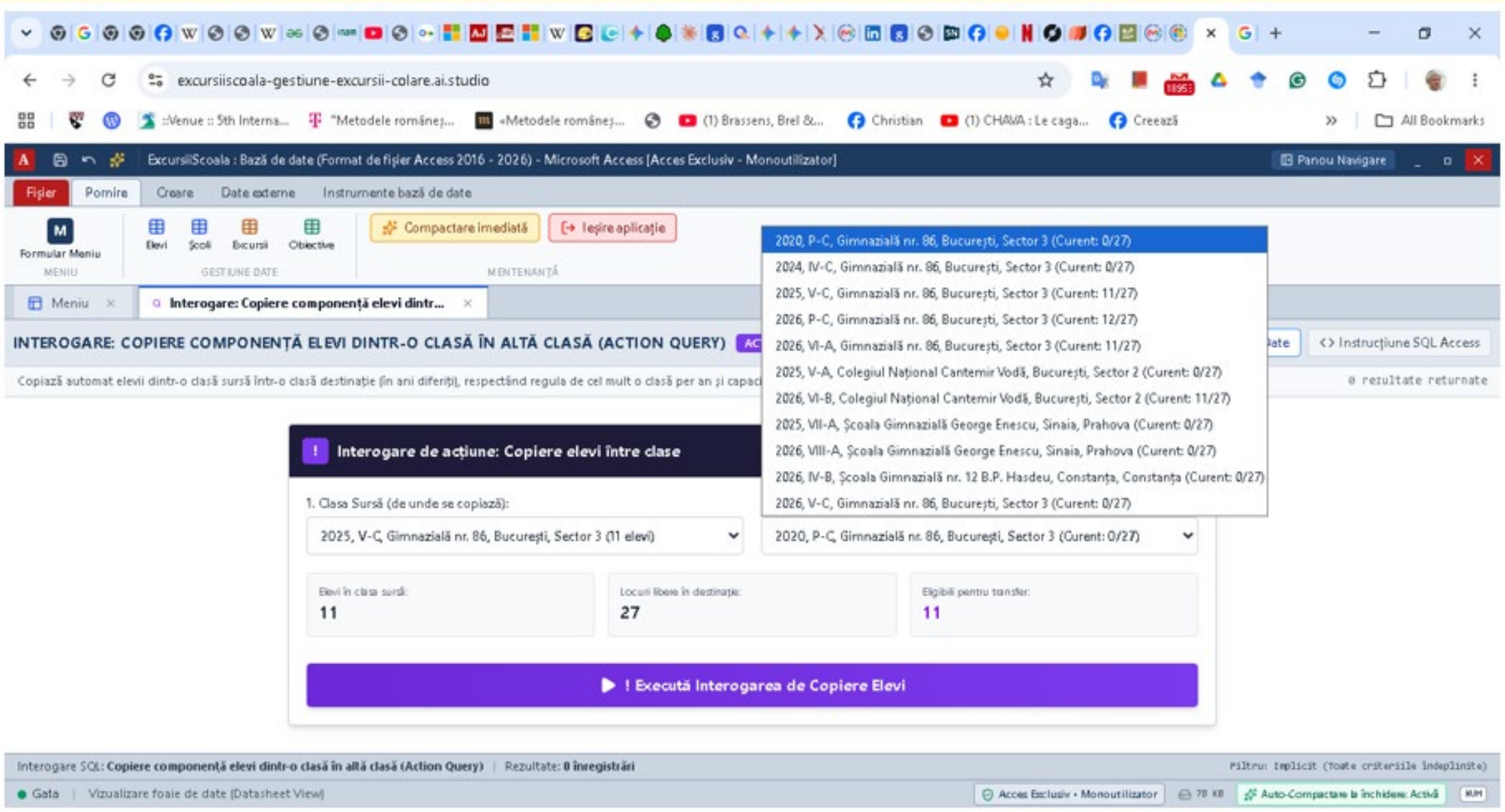

***Figure 23:*** The second action query.

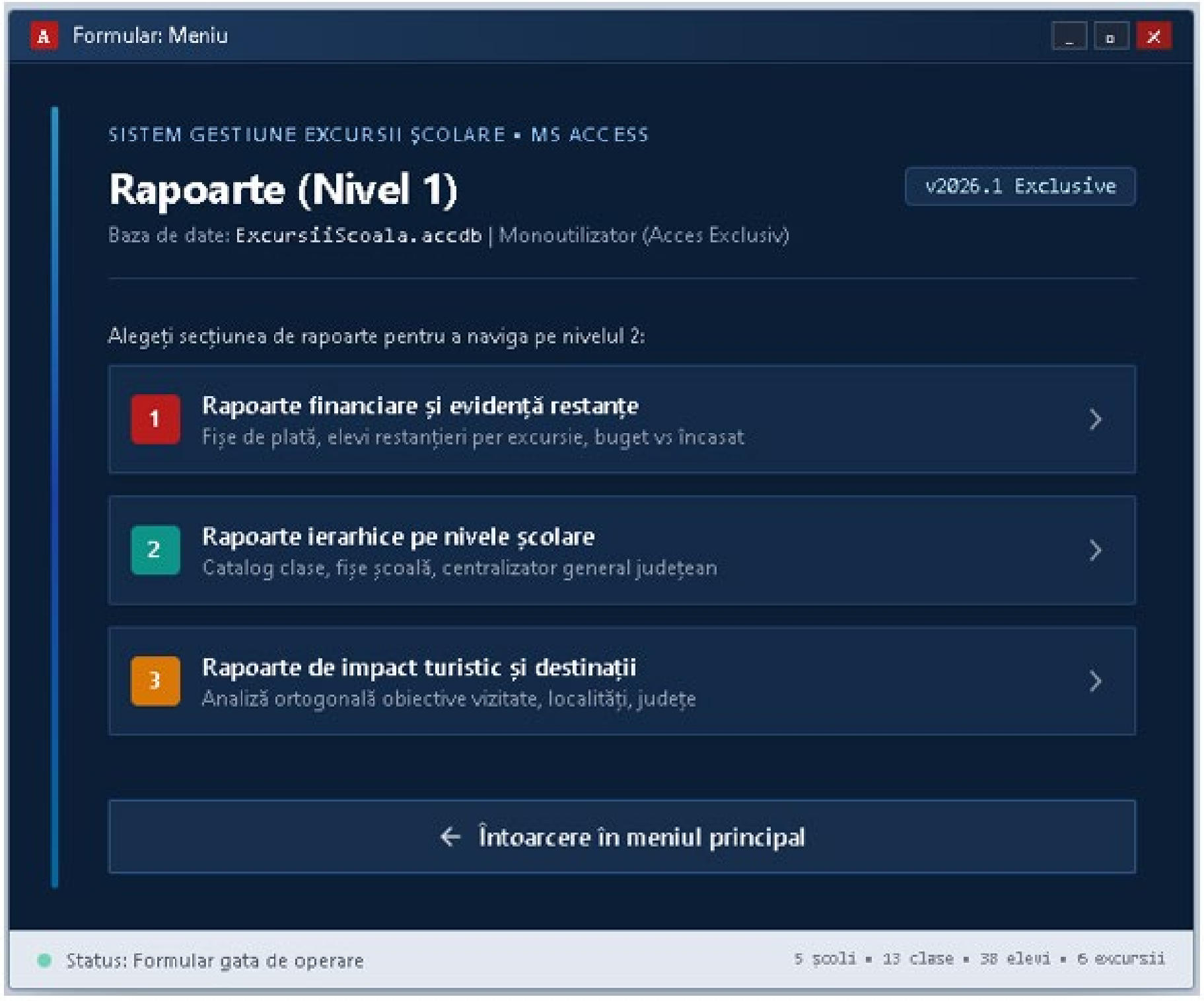

***Figure 24:*** The 3 report groups.

The first group contains the two reports corresponding to the two financial queries. The first one was presented in Figure 9. The second one (see Figure 25), corresponding to the problematic query shown in Figures 18 and 19, is syntactically ok; however, it would be more useful were it also contain a total debt column (obtained by subtracting, on every row, the value in the last column from the one in the previous column), sparing its users from computing it manually, as well as totals (per classes, schools, localities, counties, years, and general, for all stored data). Unfortunately, instead of intelligently sorting data, e.g., ascending on county name, locality name, school name, class name, and trip date this report is unsorted.

The second group contains 3 reports (see Figure 26). The first one (see Figure 27) computes the total enrolled children per school, year, and classes, commenting on its last column whether their number obeys the interval [10, 27] (in green) or not (in red). Unfortunately, instead of intelligently sorting data, e.g., first on the last column (first red ones, then green ones), then descending on years, then ascending on pupils number, county name, locality name, school name, and class name, this report too is unsorted. Moreover, it would be much more useful were it also providing totals (per schools, localities, counties, years, categories, and general, for all stored data).

The second hierarchical report lists all schools, their classes, and the total of trips organized. Again, unsorted and without group and general totals: it should be grouped per counties, localities, schools, and classes.

The third one (see Figure 28) lists the total number of localities, schools, and touristic objectives per county, plus a national total of school trips, participants, organizing classes, and involved schools. Of course, its detail data should be sorted ascendingly on county names.

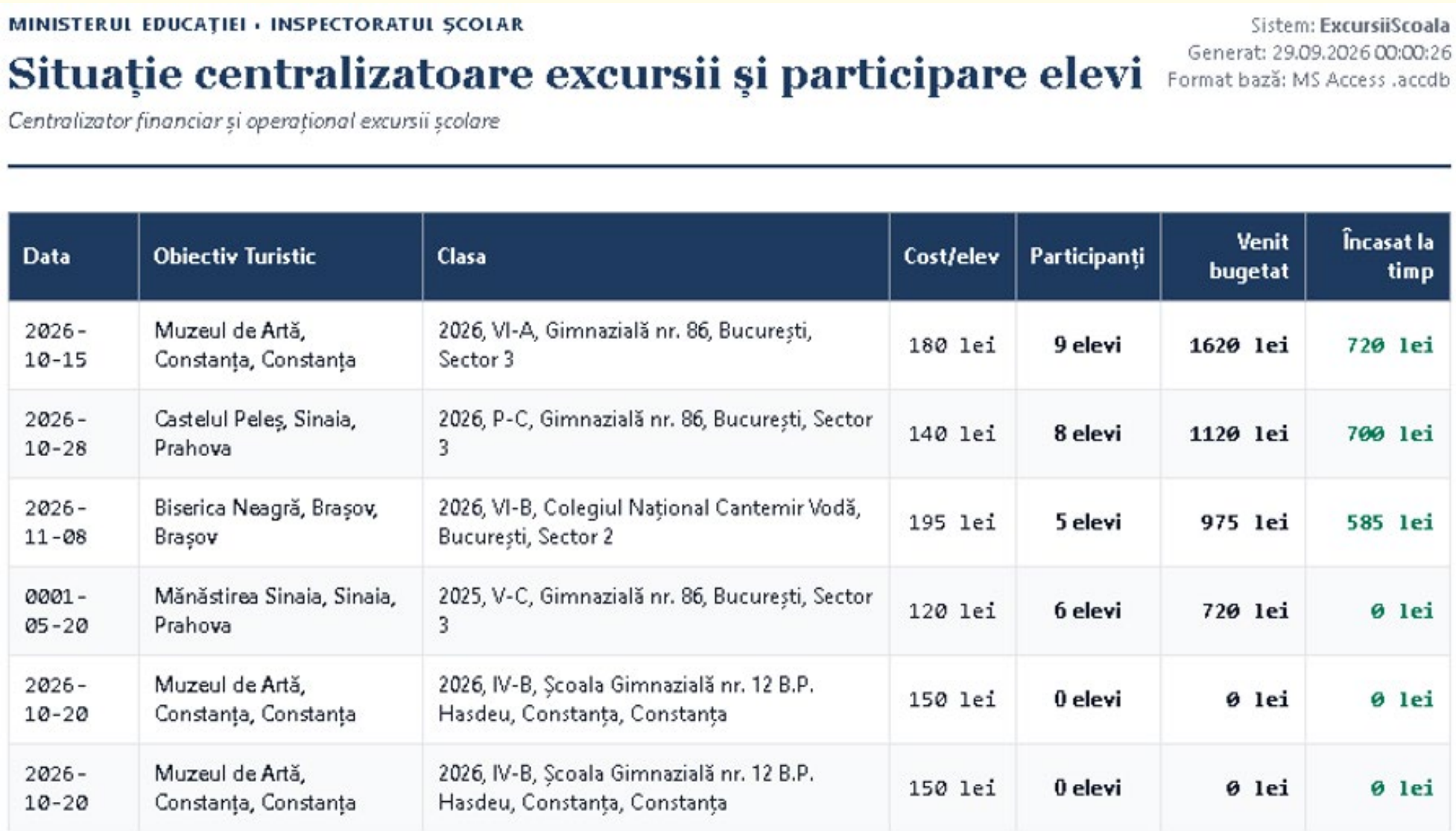

MINISTERUL EDUCAȚIEI • INSPECTORATUL ȘCOLAR

Sistem: ExcursiiScoala
Generat: 29.09.2026 00:00:26
Format bază: MS Access .accdb

**Situație centralizatoare excursii și participare elevi**

*Centralizator financiar și operațional excursii școlare*

| Data | Obiectiv Turistic | Clasa | Cost/elev | Participanți | Venit bugetat | Încasat la timp |
|---|---|---|---|---|---|---|
| 2026-10-15 | Muzeul de Artă, Constanța, Constanța | 2026, VI-A, Gimnazială nr. 86, București, Sector 3 | 180 lei | 9 elevi | 1620 lei | 720 lei |
| 2026-10-28 | Castelul Peleș, Sinaia, Prahova | 2026, P-C, Gimnazială nr. 86, București, Sector 3 | 140 lei | 8 elevi | 1120 lei | 700 lei |
| 2026-11-08 | Biserica Neagră, Brașov, Brașov | 2026, VI-B, Colegiul Național Cantemir Vodă, București, Sector 2 | 195 lei | 5 elevi | 975 lei | 585 lei |
| 0001-05-20 | Mănăstirea Sinaia, Sinaia, Prahova | 2025, V-C, Gimnazială nr. 86, București, Sector 3 | 120 lei | 6 elevi | 720 lei | 0 lei |
| 2026-10-20 | Muzeul de Artă, Constanța, Constanța | 2026, IV-B, Școala Gimnazială nr. 12 B.P. Hasdeu, Constanța, Constanța | 150 lei | 0 elevi | 0 lei | 0 lei |
| 2026-10-20 | Muzeul de Artă, Constanța, Constanța | 2026, IV-B, Școala Gimnazială nr. 12 B.P. Hasdeu, Constanța, Constanța | 150 lei | 0 elevi | 0 lei | 0 lei |

***Figure 25:*** The second financial report (corresponding to the query from Figure 18).

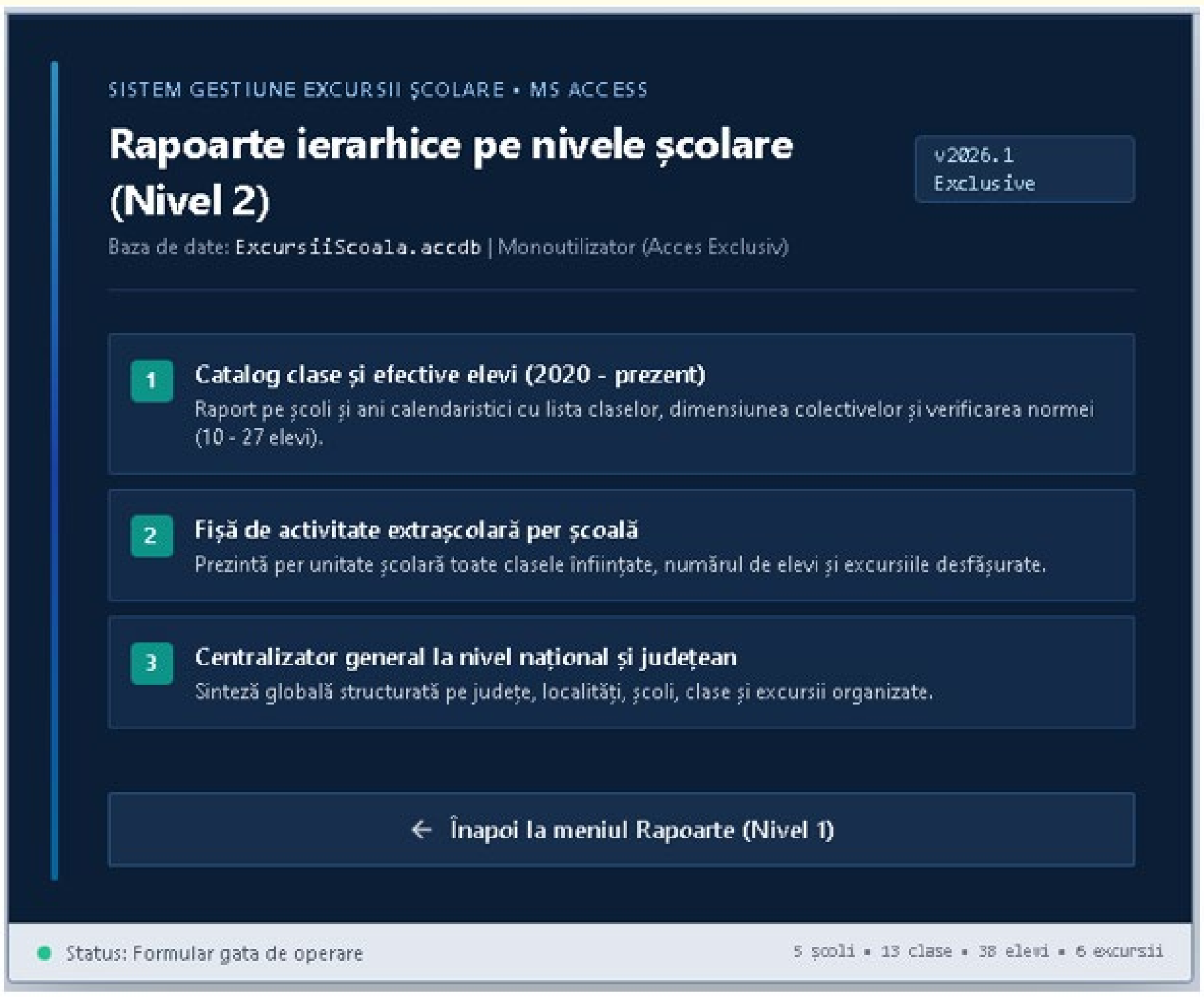

***Figure 26:*** The reports of the hierarchical group.

MINISTERUL EDUCAȚIEI • INSPECTORATUL ȘCOLAR

Sistem: ExcursiiScoala
Generat: 29.09.2026 00:27:22
Format bază: MS Access .accdb

**Catalog clase și efective elevi (2020 - prezent)**

*Evidența claselor și distribuția elevilor pe ani de studiu*

| Școala | An calendaristic | Clasa | Efectiv elevi | Statut normă legală (10 - 27 elevi) |
|---|---|---|---|---|
| Gimnazială nr. 86 | 2020 | P-C | 0 elevi | ⚠ Sub efectivul minim (0/10) |
| Gimnazială nr. 86 | 2024 | IV-C | 0 elevi | ⚠ Sub efectivul minim (0/10) |
| Gimnazială nr. 86 | 2025 | V-C | 11 elevi | Conformă regulament |
| Gimnazială nr. 86 | 2026 | P-C | 12 elevi | Conformă regulament |
| Gimnazială nr. 86 | 2026 | VI-A | 11 elevi | Conformă regulament |
| Colegiul Național Cantemir Vodă | 2025 | V-A | 0 elevi | ⚠ Sub efectivul minim (0/10) |
| Colegiul Național Cantemir Vodă | 2026 | VI-B | 11 elevi | Conformă regulament |
| Școala Gimnazială George Enescu | 2025 | VII-A | 0 elevi | ⚠ Sub efectivul minim (0/10) |
| Școala Gimnazială George Enescu | 2026 | VIII-A | 0 elevi | ⚠ Sub efectivul minim (0/10) |
| Școala Gimnazială nr. 12 B.P. Hasdeu | 2026 | IV-B | 0 elevi | ⚠ Sub efectivul minim (0/10) |
| Gimnazială nr. 86 | 2026 | V-C | 0 elevi | ⚠ Sub efectivul minim (0/10) |
| Gimnazială nr. 86 | 2025 | P-C | 25 elevi | Conformă regulament |

***Figure 27:*** The first hierarchical report.

Finally, Figure 29 shows the only report from the third group, the orthogonal one on visited touristic objectives during scholar trips: it lists the objectives, corresponding localities and counties names, their total number of visiting scholar trips, and children. Of course, it should be sorted (e.g., ascendingly on county, locality, and objective names or, more intelligently, descending on children and trip totals and then ascending on county and locality names). Moreover, the header should mention the period for which the totals were computed (in this demo data case, 2020 - 2026).

Generally, any developer would like that at least the SQL SELECT statements that compute the reports instances be also available, like it is for queries (but editable, not just read-only).

## Conclusion and Further Work

The main contributions of this paper are the proofs that even non-programmers may vibe code using Google AI Studio (which is very probable the case for other similar AI tools), and even in a foreign language other than English, that the correspondingly generated apps must be thoroughly tested, and that, while displaying lot of very useful and nice looking features, these apps may contain lot of bugs and not that intelligent ones.

Summarizing, here are their most important best features, in the order in which we discovered them: they are generated in a few minutes, their foreign language accuracy is impressive, their GUIs are of high quality, their db schema and SQL queries are almost perfectly designed and developed, their code enforces all requested business rules, their demo data is plausible, obeying all business rules, invalid data is not displayed in the combo-boxes instances, and automatic combo-box instance requery immediately after their row sources are updated.

MINISTERUL EDUCAȚIEI • INSPECTORATUL ȘCOLAR

## Centralizator general la nivel național și județean

*Bilanț statistic general ExcursiiScoala*

Sistem: **ExcursiiScoala**
Generat: 29.09.2026 11:39:46
Format bază: MS Access .accdb

| ȘCOLI PARTENERE | CLASE TOTALE | ELEVI ÎNREGISTRAȚI | EXCURSII DERULATE |
|---|---|---|---|
| 5 | 13 | 38 | 6 |

| Județ | Localități | Școli | Obiective Turistice |
|---|---|---|---|
| Sector 3 | 1 | 1 | 1 |
| Sector 1 | 1 | 0 | 0 |
| Sector 2 | 1 | 1 | 0 |
| Sector 5 | 1 | 0 | 1 |
| Prahova | 2 | 1 | 2 |
| Constanța | 2 | 1 | 2 |
| Brașov | 2 | 1 | 2 |
| Cluj | 1 | 0 | 1 |
| Sibiu | 1 | 0 | 2 |

***Figure 28:*** The third hierarchical report.

MINISTERUL EDUCAȚIEI • INSPECTORATUL ȘCOLAR

## Analiză ortogonală obiective turistice vizitate

*Clasament și distribuție geografică a obiectivelor turistice*

Sistem: **ExcursiiScoala**
Generat: 29.09.2026 11:48:06
Format bază: MS Access .accdb

| Obiectiv Turistic | Localitate & Județ | Excursii primite | Total elevi vizitatori |
|---|---|---|---|
| Muzeul de Artă | Constanța, Constanța | 3 | 9 elevi |
| Delfinariu și Acvariu | Constanța, Constanța | 0 | 0 elevi |
| Castelul Peleș | Sinaia, Prahova | 1 | 8 elevi |
| Mănăstirea Sinaia | Sinaia, Prahova | 1 | 6 elevi |
| Biserica Neagră | Brașov, Brașov | 1 | 5 elevi |
| Cetatea Râșnov | Râșnov, Brașov | 0 | 0 elevi |
| Muzeul Național de Istorie Naturală Grigore Antipa | București, Sector 3 | 0 | 0 elevi |
| Palatul Parlamentului | București, Sector 5 | 0 | 0 elevi |
| Grădina Botanică Alexandru Borza | Cluj-Napoca, Cluj | 0 | 0 elevi |
| Muzeul Național Brukenthal | Sibiu, Sibiu | 0 | 0 elevi |

***Figure 29:*** The only hierarchical report from the third group.

Dually, here are their worst unpleasant features, in the same order: you only have end-user access to a web-based app stored in the cloud, which you cannot download and install on your PC/phone, although we asked for a specific technology, MS Access, in fact possibly only the underlying used RDBMS is its db engine, while the frontend one it is not (probably it is developed in Python instead), thus loosing very end-user important features (e.g., freely resizing and changing column position in forms, sorting in any desired order of their instances, freely filtering data, vertically displaying embedded subforms, etc.), form, report, and combo-box instances always displayed unsorted and ungrouped, lacking group and general totals, data displayed without possibility to uniquely identify the corresponding objects/records (as parts of their semantic unique keys are missing), rules enforced only when using calendar controls, not also when calendar dates are entered manually, computed columns whose values are not updated after underlying values are updated, not automatically enforcing a semantic unique key on any fundamental table which does not have any such key, not automatically enforcing trivially inferable rules from the ones explicitly included in the spec doc, and automatic return to the app main menu after closing a query or report.

Finally, here are the additional features that we would like to benefit from the next Google AI Studio versions, in the same order: obtaining apps generated in exactly the technology requested, the possibility to upload the generated up locally or at least publish it with a cloud provider at your choice, saving db backups in the corresponding RDBMS format as well (not only in .json) and restoring them, full (not just read-only) access to query and report SQL statements, possibility to copy and resize displayed data in both forms and reports, to freely change column positions and sort their instances in any form, in any used frontend technology, automatically adding filtering capabilities to all forms and reports, except when explicitly forbidden in the spec docs, and never display unsorted data in either combo-boxes, forms, or reports: choose instead a context-based intelligent sorting order.

We are currently in the process of gathering and classifying all the bugs and improvement suggestions reported in this case study. We would also add our proposed general axioms and then submit them all to the AI Google Studio. Our next paper will be dedicated to the answers from it and, especially, to the test results of the new corresponding app version.

***Conflict of interest***

The authors declare that the research was conducted in the absence of any commercial or financial relationships that could be construed as a potential conflict of interest.

***Acknowledgements***

This research was not sponsored by anybody and nobody other than its authors contributed to it. The authors are grateful to Mihaela Virginia Mancas and Diana Christina Mancas, who are always carefully checking all our manuscripts.

## References

1. Sarkar A and Drosos I. "Vibe coding: programming through conversation with artificial intelligence". arXiv (2025). https://arxiv.org/abs/2506.23253
2. Coursera. "Google AI Professional Certificate" (2026). https://www.coursera.org/professional-certificates/google-ai
3. Coursera. "Specialization Certificate Google AI" (2026). https://www.coursera.org/account/accomplishments/professional-cert/1PBKSC4ZD4ZA
4. Google Corp. "Google AI Studio" (2026). https://aistudio.google.com/apps?project=gen-lang-client-0796510039
5. Google Corp. "Gemini API" (2026). https://aistudio.google.com/docs
6. Google Corp. "Gemini API developer stories" (2026). https://aistudio.google.com/case-studies
7. Google Corp. "Google DeepMind Publications" (2026). https://deepmind.google/research/publications/
8. Google AI for Developers. "Build with Google AI Forum" (2026). https://discuss.ai.google.dev
9. Anthropic Corp. "Claude. Your thinking partner for big ambitions" (2026). https://claude.ai/login
10. Anthropic Corp. "Introducing Claude Corps" (2026). https://www.anthropic.com/news/claude-corps

11. OpenAI. “ChatGPT” (2026). https://chatgpt.com/
12. Horvat M and Kralj B and Gledec G. “A Comparative Study of Vibe Coding with ChatGPT and Gemini in Front-end Web Development”. In: Redep MB and Buibanic M (eds). Proc. CECIS 2025, University of Zagreb, Croatia (2025): 787-796.
13. Base44. “Build powerful apps and websites fast” (2026). https://app.base44.com/welcome
14. Mancas C and Mancas DC. “From (Elementary) Mathematical Data Model Schemas to Safe Blazor Web Applications with Claude AI”. PriMera Scientific Engineering. 8.4 (2026): 16-37. https://doi.org/10.56831/PSEN-08-261
15. Microsoft Corp. “Access help & learning” (2026). https://support.microsoft.com/en-us/access
16. Codd EF. “A relational model for large shared data banks”. CACM 13.6 (1970): 377-387.
17. Abiteboul S, Hull R and Vianu V. “Foundations of Databases”. Addison-Wesley, Reading, MA (1995).
18. Mancas C. “Conceptual Data Modeling and Database Design: A Completely Algorithmic Approach. Volume 1: The Shortest Advisable Path”. Apple Academic Press, Waretown, NJ (2015).
19. Google AI Studio. “ExcursiiScoala” (2026). https://excursiiscoala-gestiune-excursii-colare.ai.studio/
20. Mancas C. “Algorithms for Database Keys Discovery Assistance”. In: Repa V and Bruckner T (eds). Procs. BIR 2016, LNBIP 261, Springer, Cham (2016): 322-338. https://doi.org/10.1007/978-3-319-45321-7_23